# Giant domain wall anomalous Hall effect in a layered antiferromagnet EuAl$_2$Si$_2$

Wei Xia,[1,8,†] Bo Bai,[1,†] Xuejiao Chen,[2,†] Yichen Yang,[3,†] Yang Zhang,[1] Jian Yuan,[1] Qiang Li,[4] Kunya Yang,[5] Xiangqi Liu,[1] Yang Shi,[4] Haiyang Ma,[1] Huali Yang,[2] Mingquan He,[5] Lei Li,[6] Chuanying Xi,[7] Li Pi,[7] Xiaodong Lv,[9] Xia Wang,[10] Xuerong Liu,[1] Shiyan Li,[4] Xiaodong Zhou,[11] Jianpeng Liu,[1,8] Yulin Chen,[1,8,12] Jian Shen,[4,11] Dawei Shen,[3,13,*] Zhicheng Zhong,[2,*] Wenbo Wang,[1,8,*] and Yanfeng Guo[1,8,*]

[1]School of Physical Science and Technology, ShanghaiTech University, Shanghai 201210, China

[2]CAS Key Laboratory of Magnetic Materials and Devices Zhejiang Province Key Laboratory of Magnetic Materials and Application Technology, Ningbo Institute of Materials Technology and Engineering, Chinese Academy of Sciences, Ningbo 315201, China

[3]State Key Laboratory of Functional Materials for Informatics, Shanghai Institute of Microsystem and Information Technology, Chinese Academy of Sciences, Shanghai 200050, China

[4]State Key Laboratory of Surface Physics, Department of Physics, Fudan University, Shanghai 200433, China

[5]Low Temperature Physics Lab, College of Physics; Center of Quantum Materials and Devices, Chongqing University, Chongqing 401331, China

[6]Frontiers Science Center for Flexible Electronics, Xi'an Institute of Flexible Electronics (IFE) and Xi'an Institute of Biomedical Materials & Engineering, Northwestern Polytechnical University, Xi'an 710072, China

[7]Anhui Province Key Laboratory of Condensed Matter Physics at Extreme Conditions, High Magnetic Field Laboratory of the Chinese Academy of Sciences, Hefei, Anhui 230031, China

[8]ShanghaiTech Laboratory for Topological Physics, ShanghaiTech University,

Shanghai 201210, China

[9]College of Physics and Electronic Information, Inner Mongolia Normal University, 81 Zhaowuda Road, Hohhot 010022, Inner Mongolia,

[10]Analytical Instrumentation Center, School of Physical Science and Technology, ShanghaiTech University, Shanghai 201210, China

[11]Institute for Nanoelectronic Devices and Quantum Computing, Fudan University, Shanghai 200433, China

[12]Clarendon Laboratory, Department of Physics, University of Oxford, Oxford OX1 3PU, United Kingdom

[13]National Synchrotron Radiation Laboratory, University of Science and Technology of China, 42 South Hezuohua Road, Hefei, Anhui 230029, China

**Generally, the dissipationless Hall effect in solids requires time-reversal symmetry breaking (TRSB), where TRSB induced by external magnetic field results in ordinary Hall effect, while TRSB caused by spontaneous magnetization gives rise to anomalous Hall effect (AHE) which scales with the net magnetization. The AHE is therefore not expected in antiferromagnets with vanishing small magnetization. However, large AHE was recently observed in certain antiferromagnets with noncolinear spin structure and nonvanishing Berry curvature. Here, we report another origin of AHE in a layered antiferromagnet $EuAl_2Si_2$, namely the domain wall (DW) skew scattering with Weyl points near the Fermi level, in experiments for the first time. Interestingly, the DWs form a unique periodic stripe structure with controllable periodicity by external magnetic field, which decreases nearly monotonically from 975 nm at 0 T to 232 nm at 4 T. Electrons incident on DW with topological bound states experience strong asymmetric scattering, leading to a giant AHE, with the DW Hall conductivity (DWHC) at 2 K and 1.2 T reaching a record value of $\sim 1.51 \times 10^4$ S cm$^{-1}$ among bulk systems and being two orders of magnitude larger than the intrinsic anomalous Hall conductivity. The observation**

**not only sets a new paradigm for exploration of large anomalous Hall effect, but also provides potential applications in spintronic devices.**

†These authors contributed equally to this work:

Wei Xia, Bo Bai, Xuejiao Chen and Yichen Yang.

*Correspondence:

D.W.S.(dwshen@ustc.edu.cn), Z.C.Z.(zhong@nimte.ac.cn),

W.B.W.(wangwb1@shanghaitech.edu.cn), and Y.F.G. (guoyf@shanghaitech.edu.cn)

**INTRODUCTION**

In magnets, net magnetization allows a spontaneous Hall effect even in the absence of external magnetic field, which is termed as anomalous Hall effect (AHE) that distinguishes from the ordinary Hall effect [1]. The AHE often appeared in ferromagnets usually scales with the net magnetization and is therefore not expected in conventional antiferromagnets with collinear spin arrangements and hence vanishing small magnetization. In theory, AHE is suggested to arise from a pseudo-magnetic field caused by the momentum integrated Berry curvature, which comes from the entangled Bloch electronic bands with spin-orbit coupling (SOC) and takes nonzero value in a system with time-reversal symmetry breaking (TRSB) [1-4]. Generally, TRSB in solids can be induced by external magnetic field, such as in nonmagnetic compounds including $ZrTe_5$ [5], $ScV_6Sn_6$ [6], $KV_3Sb_5$ [7] and the dilute magnetically doped Kondo systems [8], or by the presence of spontaneous net magnetization such as in ferromagnetic metals. However, in some antiferromagnets with noncolinear spin structure and nonvanishing Berry curvature, large AHE with the magnitude even comparable with that in ferromagnets was reported. The explicit examples are the hexagonal chiral antiferromagnets $Mn_3X$ (X = Ga, Ge, Sn, Pt) [9-12], in which geometry frustration in the kagome lattice of Mn atoms within the *ab*-plane, together with the Dzyaloshinskii-Moria interaction, leads to noncolinear inverse triangular spin structure with the moments aligned at 120º. This peculiar spin structure in $Mn_3Sn$ produces a remarkably large AHE with the anomalous Hall conductivity (AHC) reaching 20 S cm$^{-1}$ at room temperature and larger than 100 S cm$^{-1}$ at low temperature [9], despite of a very small magnetization of only ~ 0.002 $\mu_B$/Mn. Soon after, AHE was also discovered in several AFM topological materials such as the quantum anomalous Hall systems [13], $EuCd_2X_2$ (X = As, Sb) [14-16], and half-Heusler compounds *A*PtBi (*A* = Gd, Nd) [17, 18].

Because intrinsic Berry curvature is initially related to band structure topology and topological invariants, the success of Berry curvature physics therefore provides a general classification of band topology and eventually led to the discovery of quantum

AHE with the Hall conductivity taking a quantized value determined by the Chern number [12]. This guides to the study of topological properties through analyzing the AHE. However, seen from the illustrations of various Hall effect in Table S1 of the Supplementary Information (SI), besides intrinsic Berry curvature, extrinsic mechanisms could also generate AHE, due to the electrons scattering off of a sudden change in the periodic potential caused by structural defects or chemical and magnetic impurities [19-21]. The extrinsic mechanisms are dominated by the skew scattering which includes electrons deflecting transversely by nonmagnetic impurities [22, 23], spin-dependent electron scattering on localized magnetic moments [24], and scattering off of spin clusters [7, 25-28]. The non-coplanar spin cluster with spin chirality, in particular, was predicated to be one key ingredient towards the generation of giant skew scattering in non-centrosymmetric or frustrated magnets [28]. Recently, a domain wall (DW) skew scattering mechanism for AHE, irrelevant to spin chirality, was theoretically predicted in Weyl semimetals [29]. In sharp contrast with spin cluster skew scattering driven AHE, this effect is mainly proportional to DW density and will be significantly enhanced when the Weyl points (WPs) are close to the Fermi level $E_F$. In particular, dense parallel magnetic DW structure was proposed to generate a very strong AHE, even with the strength comparable with intrinsic AHE from bulk. Nevertheless, this type of ordered DW structure is rather rare in magnetic materials, where disordered and curved DWs are often observed [30, 31].

In this work, we present our observation of giant DW Hall effect (DWHE) with the DW Hall conductivity (DWHC) even reaching ~ $10^4$ S cm$^{-1}$ at 2 K and 1.2 T in a layered antiferromagnet EuAl$_2$Si$_2$, which is the highest among all known bulk AHE materials and is even two orders of magnitude larger than the bulk intrinsic AHC. This is the first example of so large DWHE in an antiferromagnet originated from periodic stripe DW skew scattering, which definitely would prompt future exploration of large extrinsic AHE. Furthermore, the magnetic field controllable DW density and hence magnitude of DWHE in a layered antiferromagnet offers potential applications in high density data storage and processing.

The details for crystal growth and quality examinations, magnetization, magnetotransport, angle-resolved photoemission spectroscopy, magnetic force microscopy (MFM) measurements, first-principles calculations, and detailed data analysis of bulk EuAl$_2$Si$_2$ are presented in the SI which includes references [1, 5, 8, 9, 11, 12, 15, 17, 23, 32-104].

**RESULTS AND DISCUSSION**

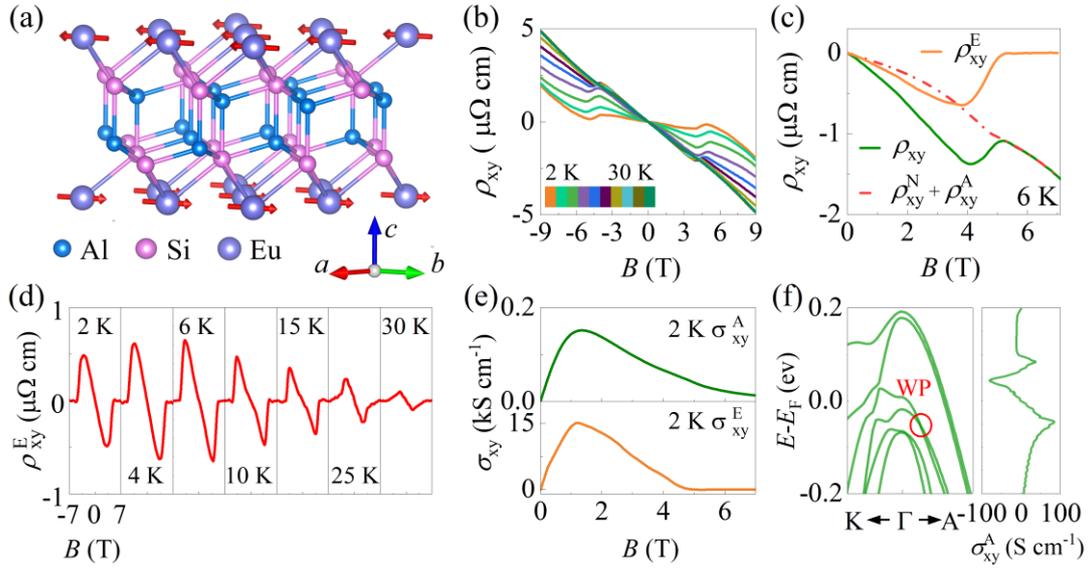

**Fig. 1.** (a) Crystal structure of EuAl$_2$Si$_2$. The arrows denote the spin directions. (b) Magnetic field $B$ dependence of transverse Hall resistivity $\rho_{xy}$ ($B//c$-axis and $B\perp I$) at various temperatures. (c) The $\rho_{xy}$ (green) measured at 6 K, the calculated $R_H B + 4\pi R_S M$ fitting results (red) and extracted $\rho^E_{xy}$ (orange). (d) Magnetic field dependence of $\rho^E_{xy}$ at various temperatures. (e) Top: Intrinsic AHC $\sigma^A_{xy}$ at 2 K. Bottom: Extracted Hall conductivity $\sigma^E_{xy}$ at 2 K. (f) Left: Calculated band structure along the K-Γ-A direction of the Brillouin zone. The red circle marks the WPs. Right: The calculated AHC $\sigma^A_{xy}$ of the WPs.

EuAl$_2$Si$_2$ crystallizes into the CaAl$_2$Si$_2$-type trigonal structure with the space group $P\bar{3}m1$ (No. 164). As shown in Fig. 1(a), the unit cell consists of two Al-Si zigzag chains which are interlaced and superimposed along the *c*-axis between two Eu atom layers. It exhibits an antiferromagnetic (AFM) order at the Néel temperature $T_N \sim 33.6$ K, and Eu atoms form an in-plane A-type AFM structure below $T_N$ [34, 56, 105, 106], with spin directions along the *b*-axis, as denoted by the arrows illustrated in Fig. 1(a). The magnetic susceptibility is isotropic in the paramagnetic region, while exhibits clear anisotropy in the ordered state. The magnetic phase diagram established based on ac magnetic susceptibility measurements, as shown in Fig. S2(d) of the SI, exposes that an external magnetic field of about 4 T can fully polarize all Eu spins along the *c*-axis into a ferromagnetic (FM) state, which is signified by a clear anomaly around 4 T in the transverse Hall resistivity $\rho_{xy}$ in Fig. 1(b). In general, $\rho_{xy}$ in a magnetic topological material is expressed as $\rho_{xy} = \rho^N_{xy} + \rho^A_{xy} + \rho^E_{xy} = R_H B + 4\pi R_S M + \rho^E_{xy}$, where $\rho^N_{xy}$ comes from the ordinary Hall effect, $\rho^A_{xy}$ denotes intrinsic AHE caused by nonvanishing Berry curvature and $\rho^E_{xy}$ is arisen from other extrinsic sources [32, 93]. It is apparent that $\rho^E_{xy}$ can be obtained by subtracting the ($\rho^N_{xy} + \rho^A_{xy}$) item. To guide a clear comparison, $\rho_{xy}$, $\rho^N_{xy} + \rho^A_{xy}$ and $\rho^E_{xy}$ versus $B$ at 6 K are plotted together in Fig. 1(c), which demonstrate a nonzero $\rho^E_{xy}$. The $\rho^E_{xy}$ is then individually plotted in Fig. 1(d) at different temperatures to enlarge more details, which shows the maximum value of 0.647 μΩ cm at ± 4 T and 6 K. The AHC $\sigma^A_{xy}$ [= $-\rho^A_{xy}/(\rho^2_{xy} + \rho^2_{xx})$] and $\sigma^E_{xy}$ [= $-\rho^E_{xy}/(\rho^2_{xy} + \rho^2_{xx})$] at 2 K peak around 1.2 T with the values of 151 S cm$^{-1}$ and 1.51 × 10$^4$ S cm$^{-1}$, respectively, as shown in Fig. 1(e). It is rather amazing that $\sigma^E_{xy}$ is even two orders of magnitude larger than the intrinsic $\sigma^A_{xy}$, which has never been observed in other solids, thus setting EuAl$_2$Si$_2$ a model system with giant extrinsic AHE.

To uncover the origin of giant AHE in EuAl$_2$Si$_2$, we combined first-principles calculations, magnetotransport and angle-resolved photoemission spectroscopy (ARPES) measurements to investigate the electronic band structure of both AFM and spin-polarized states. The first-principles calculations presented in the SI indicate an in-plane A-type AFM spin structure as the magnetic ground state, which is consistent with our isothermal magnetization measurements as well as with the previous neutron diffraction characterizations [56]. The magnetic space group of this A-type AFM

EuAl$_2$Si$_2$ belongs to type-IV $C_c2/c$ with BNS (Belov-Neronova-Smirnova notation) setting, which keeps inversion symmetry but breaks both $C_{3z}$ and time-reversal symmetries. Given these symmetry operators, it can still hold Z$_4$ topological numbers based on parity values [57, 58] and the calculations give Z$_4$ = 2, suggesting an axion insulator similar as AFM EuCd$_2$X$_2$ [13, 14]. In spin-polarized state, the system owns type-III $P\bar{3}m'1$ magnetic space group. For such case, the system has a TRSB while preserves spatial inversion symmetry. The calculations unveil a pair of WPs along the Γ-A direction of the Brillouin zone formed by the linear crossing of several top valence bands (see more details in SI). The WPs locate at ~ 57 meV below E$_F$, as indicated by the red circle in Fig. 1(f). The intrinsic AHC originating from the amplified Berry curvature around the WPs can be calculated, yielding a value of 98 S cm$^{-1}$ that is rather close to the experimental $\sigma^A_{xy}$ but apparently much smaller than $\sigma^E_{xy}$. Therefore, it is reasonable to exclude the amplified Berry curvature as source for the large $\sigma^E_{xy}$.

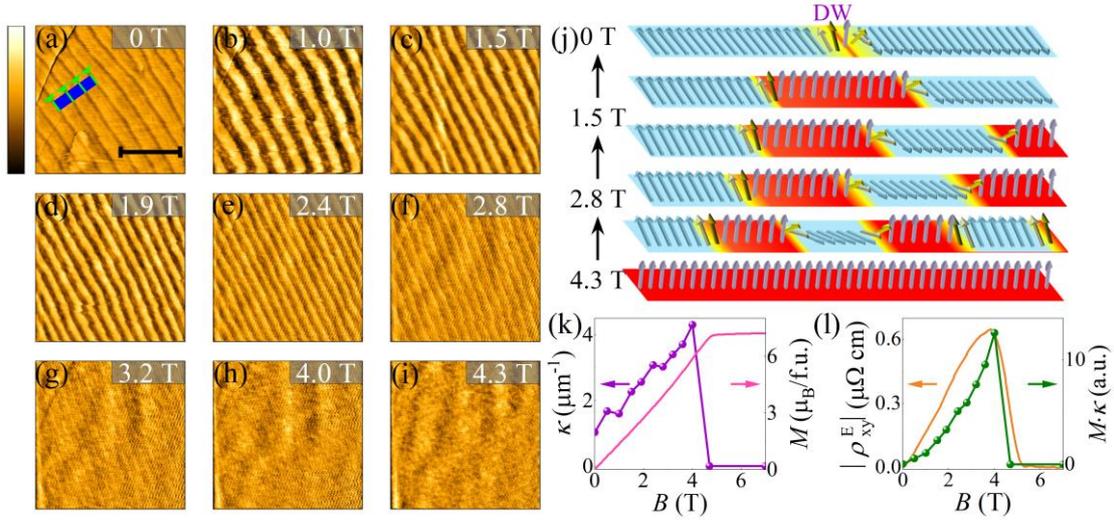

**Fig. 2.** (a)-(i) Sequential MFM images taken at 6 K under various magnetic fields from 0 to 4.3 T perpendicular to the *ab*-plane. The color scales in images (a-c) are 140 mHz, in image (d) is 80 mHz, in images (e-f) are 60 mHz, and in images (g-i) are 40 mHz. The scale bars attached to (a-i) are both 4 μm. (j) Schematic picture of domain and DWs under different magnetic fields. (k) Left: magnetic field dependence of DWD at 6 K. Right: Magnetization versus magnetic field at 6 K. (l) Left: |$\rho^E_{xy}$| versus magnetic field at 6 K. Right: $M \cdot \kappa$ versus magnetic field at 6 K.

Since the intrinsic Berry curvature cannot account for the large AHC, we then performed magnetic field dependent MFM measurements on $EuAl_2Si_2$ crystal to trace its true origin. The measurements were conducted on well cleaved *ab*-plane of the crystals at 6 K. In Fig. 2(a), interestingly, the zero magnetic field MFM image shows stripe domain structure in the A-type AFM ground state, where the bright areas refer to magnetic domains with in-plane magnetization from the surface layer. These magnetic domains are separated by dark straight lines which are AFM DWs with upward net magnetization. The MFM images were taken in the same area with increasing magnetic fields up to 4.3 T, as shown in Figs. 2(a)-2(i). As magnetic field increases, the stripe domains gradually become denser, indicating that the number of domains and DWs per unit area both increase. The corresponding spin configurations are schematically depicted in Fig. 2(j). By using Fast-Fourier transform (FFT), the extracted DWD (DW per unit length) $\kappa$ is presented in the left of Fig. 2(k). As magnetic field increases from 0 T to 1 T, the DW with upward magnetization expands and forms upward domains, as shown in Fig. 2(j). The number of DWs is doubled during this process in the fifty-fifty domain state, as depicted in Fig. 2(j). Further increasing the magnetic field leads to the enhancement of DWD. This could result from the competition between DW energy and stray field energy. Strong magnetization ($Eu^{2+}$: ~7 $\mu_B$/f.u.) and relatively weak anisotropy in materials tend to form denser domains to lower the stray field energy on the sample surface. At the same time, the MFM contrast signal significantly decreases because spins are gradually aligned to the out-of-plane direction. The DWD peaks around 4 T and the DWs become invisible with further increase of magnetic field, because the Eu spins are then fully polarized as indicated by the magnetization data in Fig. 2(k). The DWHE is proportional to DWD assuming that the contribution of every DW is identical. However, the DWs with upward and downward net magnetization have opposite contributions. For instance, adjacent DWs along the *c*-axis are coupled antiferromagnetically at the ground state, contributing to opposite skew scattering effect, as shown by Fig. S17 in the SI. Thus, DWHE is scaled with net magnetization of overall

DWs ($M_{DW}$). For simplicity, we assume that the magnetic field dependent magnetization of DWs is not far away from the bulk. Ultimately, the DWHE should be proportional to $M·κ$. The relation between $|ρ^E_{xy}|$ and $M·κ$ with increasing magnetic field is shown in Fig. 2(l). These two curves apparently follow the similar magnetic field dependent tendency, so the extra Hall resistivity $|ρ^E_{xy}|$ is very likely to be related to the DWs.

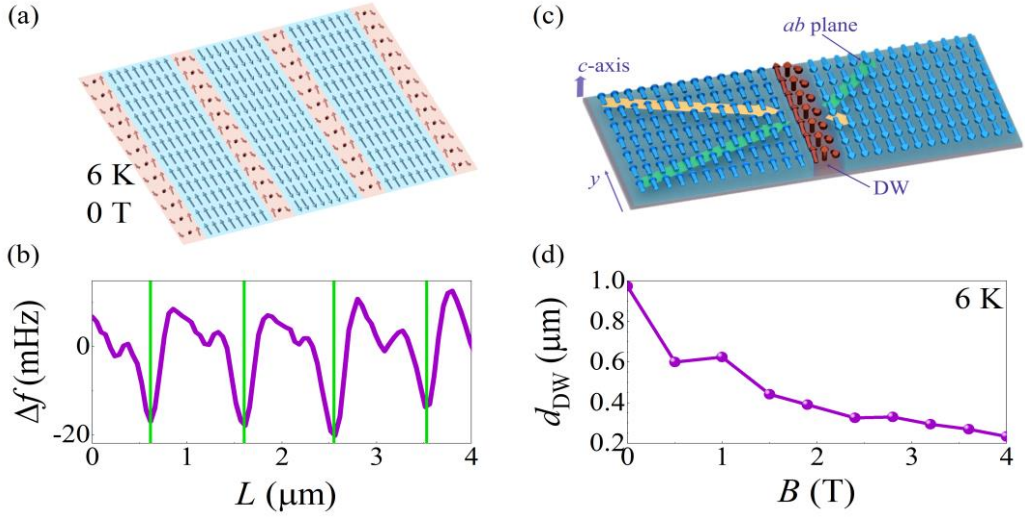

**Fig. 3.** (a) Schematic spin configuration at 6 K and 0 T. (b) A line profile of MFM signal along the blue line in Fig. 2(a) at 6 K and 0 T. Vertical green lines note the position of the DWs as indicated by the green arrows. (c) A pair of incident Bloch waves moving opposite to each other in the y-direction get transmitted asymmetrically at the DW. (d) The average distance between adjacent DW in different magnetic fields.

Fig. 3(a) shows the schematic diagram of spin configuration along a blue line drawn in Fig. 2(a), where the green arrows indicate the DWs. The MFM signal along the blue line drawn in Fig. 2(a) suggests that the average distance between two neighboring DWs is approximately 0.975 μm, as shown Fig. 3(b), which surprisingly reveals that the DW structure is in fact periodic. The valley profile of the MFM signal indicates the net magnetization at DW is parallel to *c*-axis. Fig. 3(c) shows that two incident Bloch states which move with opposite group velocities along the y-direction and get transmitted asymmetrically at the DW [29]. Since the system with DWs can be described by inhomogeneous problem, the scattering states are important for studying the transport across the DW. For such a system, the amplitude of scattering states could be expressed as a linear combination of the amplitudes of Bloch states. The coefficients of the linear combination relation can be identified with the reflection coefficients (RCs) and transmission coefficients (TCs) of the incident mode upon scattering at the DW. When the DW is in the *yz*-plane and electrons move along $k_x$ direction, due to the large asymmetrical TC along $k_y$ direction at DW, the Bloch waves get transmitted asymmetrically at the DW, thus resulting in a large Hall effect [29]. This produces a transverse Hall current due to the DW scattering when a bias voltage between the two domains is applied. A strong skew scattering will exist at the DW when the $E_F$ resides near the WPs, which suggests that the DW skew scattering will contribute additional Hall conductivity $\sigma^{DW}_{xy}$ [29, 107]. This physical picture is valid at the limit where the electron mean free path is much smaller than the linear dimensions of the domains [29], i.e. the mean free path $l_0 \ll d_{DW}$, where $d_{DW}$ is the average distance between two neighboring DWs. When *B*//*c*-axis, two Fermi pockets are resolved by quantum oscillations, and the Fermi velocities $v_F$ of these two pockets are estimated to be $2.5 \times 10^5$ and $3.8 \times 10^5$ ms$^{-1}$ (see Fig. S5 and Table S3), respectively, and the quantum scattering lifetime $\tau_Q$ are $7.0 \times 10^{-14}$ and $8.4 \times 10^{-14}$ s, respectively. The mean free path $l_0$ for $F_1$ and $F_2$ are estimated to be 17.5 nm and 31.9 nm respectively from $l_0 = \tau_Q \cdot v_F$. Fig. 3(d) shows $d_{DW}$ in different magnetic fields, which keeps the periodic feature but decreases nearly monotonically from 975 nm at 0 T to 232 nm at 4 T. As the magnetic field increases, $d_{DW}$ nearly monotonously decreases from 975 nm to 232 nm. The minimum value is also apparently larger than $l_0$. As mentioned above, the scattering between electrons and DWs will lead to strong skew scattering in transverse direction and hence an additional Hall conductivity. Our results suggest again that the large AHE observed in EuAl$_2$Si$_2$ during increase of magnetic field is very likely due to DW skew scattering.

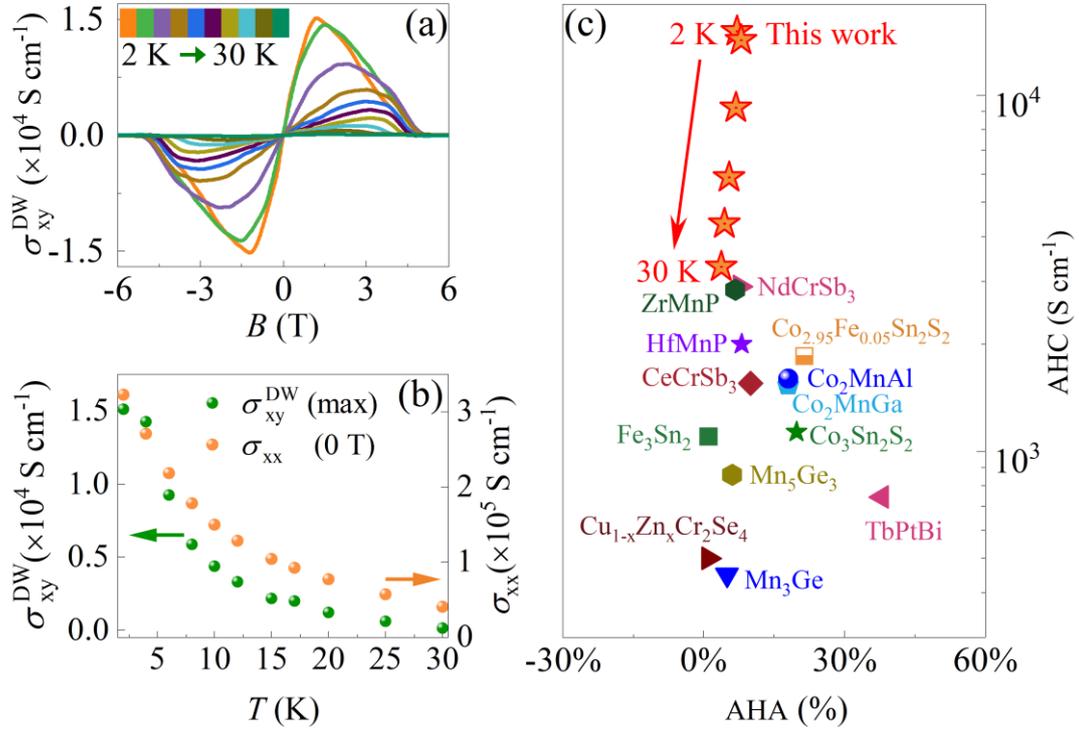

**Fig. 4.** (a) Magnetic field dependent DWs skew scattering Hall conductivity $\sigma^{DW}_{xy}$ at different temperatures. (b) Left: The maximum $\sigma^{DW}_{xy}$ of $EuAl_2Si_2$ at different temperatures. Right: The maximum $\sigma_{xx}$ of $EuAl_2Si_2$ at different temperatures. (c) A comparison with previously reported AHC values for other bulk AHE materials. The reported values were taken from references [11, 17, 23, 77-83, 108-110]. The pentagram represents the $\sigma^{DW}_{xy}$ at different temperatures in the work.

The magnetic field dependent $\sigma^{DW}_{xy}$ at different temperatures are depicted in Figs. 4(a) and 4(b), showing that the maximum values of both $\sigma^{DW}_{xy}$ and $\sigma_{xx}$ decrease as temperature increases. $\sigma^{DW}_{xy}$ reaches an extremely large value of $1.51 \times 10^4$ S cm$^{-1}$ at 2 K, as shown in Fig. 4(a), which is two orders of magnitude larger than the intrinsic AHC from bulk. In Fig. 4(c), we show a comparison of our $\sigma^{DW}_{xy}$ with previously reported AHC values of other systems with $\sigma^{A}_{xy}$ larger than 450 S cm$^{-1}$, where the data are taken from refs. [11, 17, 23, 77-83, 108-110] The reported data are completely summarized in Table S5 of SI. Apparently, the $\sigma^{DW}_{xy}$ of EuAl$_2$Si$_2$ is the largest among all reported AHC values of bulk systems. Moreover, most of the AHE systems summarized in Fig. 4(c) are FM transition metals or alloys in which the AHC is basically contributed by nontrivial Berry curvature. Within our best knowledge, only CrTe$_2$ thin flake (hundreds of nm) has a larger AHC value (~ 6.5 × 10$^4$ S cm$^{-1}$ at 2 K), but it is FM thin flake rather than bulk antiferromagnet [85]. Our observation of so large $\sigma^{DW}_{xy}$ in AFM EuAl$_2$Si$_2$ provides a paradigm of very large AHE caused by a new DW skew scattering mechanism.

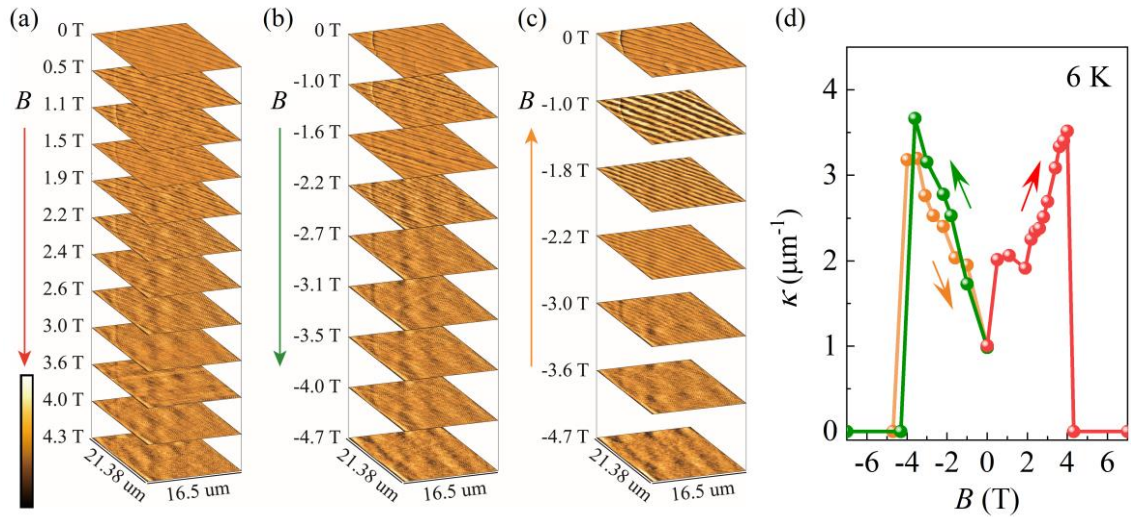

**Fig. 5.** MFM images were taken at 6 K on the same location of EuAl$_2$Si$_2$ with (a) increasing magnetic fields of 0 T to 4.3 T, (b) increasing magnetic fields of 0 T to -4.7 T and (c) decreasing magnetic fields of -4.7 T to 0 T. The color scales in images (a) are 140 mHz (0 to 1.1 T), 60 mHz (1.9 T), 40 mHz (2.2 to 4.3 T), in image (b) are 140 mHz (0 to -1.6 T), 40 mHz (-2.2 to -4.7 T), in images (c) are 140 mHz (0 to -1 T), 100 mHz (-1.8 to -2.2 T), 40 mHz (-2.2 to -4.7 T). (d) The magnetic field dependence of MFM signal frequency in (a), (b) and (c).

Figs. 5(a)-5(c) depicts positive and negative magnetic field dependent MFM images taken at 6 K on the same locations of EuAl$_2$Si$_2$. The images with increasing positive and negative magnetic fields between -4.7 T and 4.3 T show nearly similar profiles, implying symmetric DWD with respect to positive and negative magnetic

fields, which is more clearly seen by the symmetric $\kappa$ presented in Figs. 5(d). Furthermore, with decreasing negative magnetic field from -4.3 T to 0 T, the MFM images are nearly reproduced from those taken with each corresponding increasing negative magnetic field, thus exposing a reversible behavior of the DWD. This reversible behavior of the periodic stripe DWs with magnetic field makes $EuAl_2Si_2$ particularly useful for spintronic devices.

## SUMMARY

To summarize, we discovered giant anomalous Hall effect in layered antiferromagnet $EuAl_2Si_2$ with real-space spin texture. The extrinsic anomalous Hall conductivity reaches as high as $1.51 \times 10^4$ S cm$^{-1}$ at 2 K and 1.2 T, which is much larger than the intrinsic anomalous Hall conductivity and is the largest among all known bulk anomalous Hall quantum materials so far. We visualized the peculiar periodic stripe domain wall structure and its evolution against external magnetic field by using MFM measurements. We established the picture that giant anomalous Hall effect could appear due to the strong domain wall skew scattering when the Weyl points are close to the Fermi level. Furthermore, we also show that the stripe domain wall has controllable periodicity which varies nearly reversibly with increasing and decreasing magnetic field. Considering that antiferromagnets have virtues of insensitivity to perturbations, fast spin dynamics, high-density memory integration and fast data processing, and $EuAl_2Si_2$ has a layered structure that might be easily exfoliated into different layers, the observation of giant anomalous Hall effect in $EuAl_2Si_2$ would pave a new avenue for exploration of large anomalous Hall effect and offer potential applications in next-generation spintronic devices.

## ACKNOWLEDEMENTS

The authors acknowledge the National Key R&D Program of China (Grants No. 2023YFA1406100, 2022YFA1403000, and 2021YFA0718900), the Shanghai Science and Technology Innovation Action Plan (Grant No. 21JC1402000) and the National

Nature Science Foundation of China (Grants No. 920651, 11934017). Y.F.G. acknowledges the open research fund of Beijing National Laboratory for Condensed Matter Physics (2023BNLCMPKF002). W.X. thanks the support by the open project from State Key Laboratory of Surface Physics and Department of Physics, Fudan University (Grant No. KF2022_13), the Shanghai Sailing Program (23YF1426900) and the National Natural Science Foundation of China (Grants No. 12404186). W.B.W. is supported by the Science and Technology Commission of Shanghai Municipality (Grant No. 21PJ410800). D.W.S. thanks the support by National Science Foundation of China (Grant No. U2032208). J.P.L. acknowledges the Science and Technology Commission of the Shanghai Municipality (Grant No. 21JC1405100). M.Q.H. acknowledges the support by Chinesisch-Deutsche Mobilitäts program of Chinesisch-Deutsche Zentrum für Wissenschaftsförderung (Grant No. M-0496). The work at Fudan University is supported by National Natural Science Foundation of China (Grant Nos. 12074080 and 12274088), and National Key Research Program of China (2022YFA1403300). The authors also thank the support from Users with Excellence Project of Hefei Science Center CAS (No.2021HSC-UE007), Analytical Instrumentation Center (#SPST-AIC10112914) and the Double First-Class Initiative Fund of ShanghaiTech University. The authors thank Prof. Yan Sun and Prof. Andrew Boothroyd for helpful discussions.

# SI

## 1. Summary of various Hall effect

**Table S1** An illustration of anomalous Hall effect (AHE) and topological Hall effect (THE).

| | | | Origin | Illustration of the main mechanisms that can give rise to an AHE | Ref |
|---|---|---|---|---|---|
| AHE | Intrinsic defection | / | Electrons have an anomalous velocity perpendicular to the electric field related to their Berry's phase curvature | 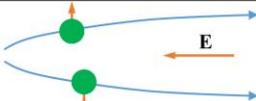 | 1 |
| | Extrinsic scattering | Side jumping | The electron velocity is deflected in opposite directions by the opposite electric fields experienced upon approaching and leaving an impurity | 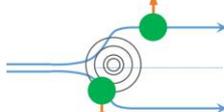 | |
| | | Skew scattering | Asymmetric scattering due to the effective spin-orbit coupling of the electron or the impurity | 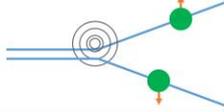 | |
| | | DW scattering | A strong skew scattering at the DW that leads to a significant additional Hall effect, When the Fermi level lies near Weyl points | 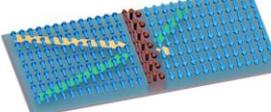 | This work |
| THE | The origin of the topological Hall effect is a Berry phase collected by the conduction electrons when following adiabatically the spin polarization of topologically stable knots in the spin structure | | | 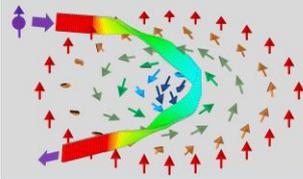 | 2, 3 |

## 2. Single crystal growth

Single crystals of EuAl$_2$Si$_2$ were grown by using the self-flux method. Starting materials of Eu ingot (99.95%), Al block (99.99%) and Si granules (99.99%) were mixed in a molar ratio of 1: 20: 4 and placed into an alumina crucible. The crucible was sealed in a quartz ampoule under vacuum. The assembly was heated in a furnace up to 1100 ℃ within 15 hours, kept at that temperature for 20 hours, and then slowly cooled down to 800 ℃ at a decreasing rate of 1.5 ℃/h. The excess flux was removed at this temperature by quickly placing the assembly into a high-speed centrifuge. Then the quartz tube was cooled down to room temperature in air, leaving EuAl$_2$Si$_2$ single crystals with black shiny metallic luster in the crucible.

## 3. Structural chracterizations

### 3.1 Single crystal X-ray diffraction

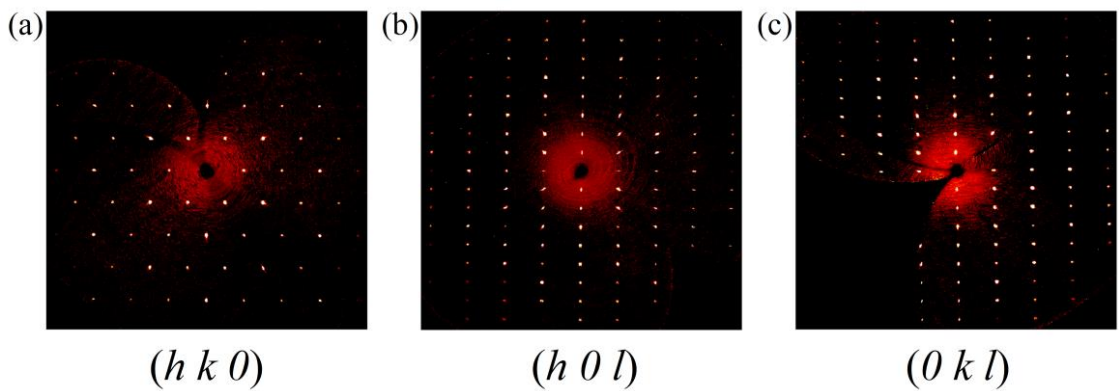

**Fig. S1.** (a), (b) and (c) show the single crystal X-ray diffraction patterns at 298 K in the reciprocal space along the (*h k 0*), (*h 0 l*) and (*0 k l*) directions.

**Table S2.** Crystallographic data of EuAl$_2$Si$_2$ summarized based on the SXRD analysis.

| Atom | Site | x | y | z | U |
|------|------|----------|----------|----------|--------|
| Eu   | 1b   | 0.000000 | 1.000000 | 0.500000 | 0.0181 |
| Al   | 2d   | 0.333333 | 0.666667 | 0.228900 | 0.0189 |
| Si   | 2d   | 0.666667 | 0.333333 | 0.129900 | 0.0207 |

Note: here 'x', 'y' and 'z' represent the normalized positions of the atoms in hexagonal crystal lattice of EuAl$_2$Si$_2$.

Single crystal X-ray diffraction was performed on a Bruker D8 single crystal X-ray diffractometer (SXRD) with Mo $K_{\alpha 1}$ ($\lambda = 0.71073$ Å) at 298 K. The pattern could be well indexed on the basis of a trigonal structure with lattice parameters $a = b = 4.178$ Å, $c = 7.249$ Å, $\alpha = 90°$, $\beta = 90°$, and $\gamma = 120°$, consistent with those reported previously [34]. The perfect reciprocal space lattice presented in Fig. S1 does not show any other miscellaneous points, indicating pure phase and high quality crystals used in this study. Results of the structure analysis based on the SXRD are summarized in Table S2.

### 3.2 Powder X-ray diffraction

The powder X-ray diffraction was performed on the *ab*-plane of EuAl$_2$Si$_2$ single crystals on a Bruker powder X-ray diffractometer with Cu $K_\alpha$ ($\lambda$ = 1.5418 Å) at 298 K. As seen in Fig. S2(a), only sharp (00*l*) Bragg diffraction peaks are visible, indicating the high quality again of our crystals. The typical picture of a crystal is presented by the inset of Fig. S2(a).

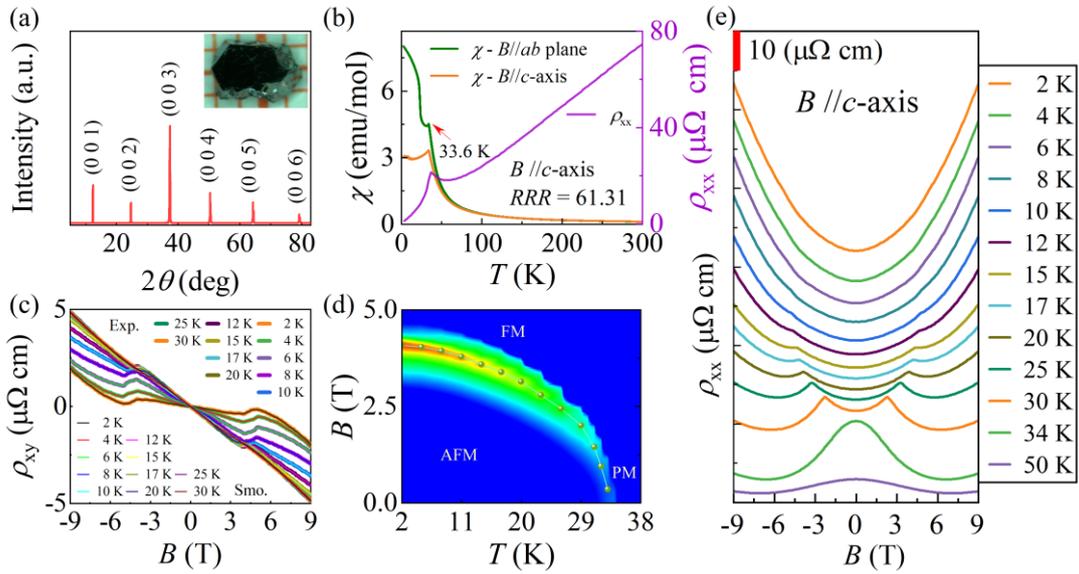

**Fig. S2.** (a) The room temperature powder X-ray diffraction pattern of EuAl$_2$Si$_2$ single crystals. Inset shows the picture of a typical single crystal. (b) Left axis: Temperature dependent magnetic susceptibility $\chi$(T). Right axis: Temperature dependent $\rho$(T) at $B$ = 0 T and between 2 - 300 K. (c) Magnetic field dependent Hall resistivity $\rho_{xy}$ (*B*//*c*-axis and $B\perp I$) at various temperatures. Thick lines represent experimental data and thin lines represent smoothed data. (d) The magnetic phase diagram of EuAl$_2$Si$_2$. The dotted line represents the critical magnetic fields at various temperatures. (e) $\rho_{xx}$ versus *B* at various temperature with *B* along the *c*-axis. The curves are shifted by offsets of $\Delta\rho_{xx}$ (5 to 15 μΩ cm).

## 4. Magnetotransport and magnetizations

The magnetotransport measurements, including the electrical resistivity and Hall effect measurements, were carried out using a standard six-probe method in a commercial DynaCool physical properties measurement system (PPMS) from Quantum Design. High magnetic field measurements were performed on the Steady High Magnetic Field Facilities installed in High Magnetic Field Laboratory, Chinese Academy of Sciences in Hefei. The magnetic susceptibility and isothermal magnetizations chracterizations were undertaken in a commercial magnetic property measurement system (MPMS-3) from Quantum Design. The ac magnetic susceptibility was measured at a magnetic field $B$ of 10 Oe and the frequency of 311 Hz between 2 and 38 K.

We now discuss the longitudinal resistivity $\rho_{xx}$ of EuAl$_2$Si$_2$ at low magnetic field between -9 T and 9 T. The $\rho_{xx}$ at zero magnetic field displays a metallic conduction behavior upon cooling with a residual resistance ratio $RRR$ (= $\rho_{xx}$(300 K)/$\rho_{xx}$(2 K)) of approximately 61.31. To eliminate the influence from Hall resistivity $\rho_{xy}$, we symmetrized the data by using the formula of $\rho_{xx}$ = [$\rho_{xx}$(+$B$) + $\rho_{xx}$(-$B$)]/2. The magnetoresistance (MR) is defined as MR = [$\rho_{xx}$($B$) - $\rho_{xx}$(0)]/$\rho_{xx}$(0) × 100%, in which $\rho_{xx}$($B$) and $\rho_{xx}$(0) represent the resistivity with and without $B$, respectively. As shown in right axis of Fig. S2(b), when $B$//$c$-axis, $\rho_{xx}$ shows a peak at the Néel temperature $T_N$ = 33.6 K, coinciding well with the antiferromagnetic (AFM) ordering temperature as seen in magnetic susceptibility $\chi$(T) measured with both $B$//$ab$-plane and $c$-axis shown in left axis of Fig. S2(b). The value is also consistent with that reported previously [34]. As shown in Fig. S2(c), we performed Hall resistivity $\rho_{xy}$ measurements with the external magnetic field applied along $c$-axis direction. The thick lines represent experimental data and the thin lines represent smoothed data. We used smoothed data to analyze $\rho_{xy}$. As seen in Fig. S2(d) by the magnetic phase diagram established on the basis of the ac magnetic susceptibility measurements, the AFM ground state can be converted into a ferromagnetic one along the $c$-axis (FM-c) by the application of vertical magnetic field of approximately 4 T. The magnetic phase transition is also reflected by the magnetic field dependent $\rho_{xx}$ presented in Fig. S2(e), which shows complicated behaviors which are nicely consistent with the magnetic phase diagram. From 50 K to 34 K, paramagnetic EuAl$_2$Si$_2$ shows metallic conduction behavior. When the temperature further decreases, a local maximum is visible in $\rho_{xx}$, which corresponds to the saturation magnetic field ($B_{sat}$) that can fully polarize the Eu spins along the $c$-axis, then the slope of $\rho_{xx}$ gradually changes its sign from negative to positive below 30 K and $B > B_{sat}$, suggesting dominant contribution of $\rho_{xx}$ from the Lorenz force induced by magnetic field [35, 36].

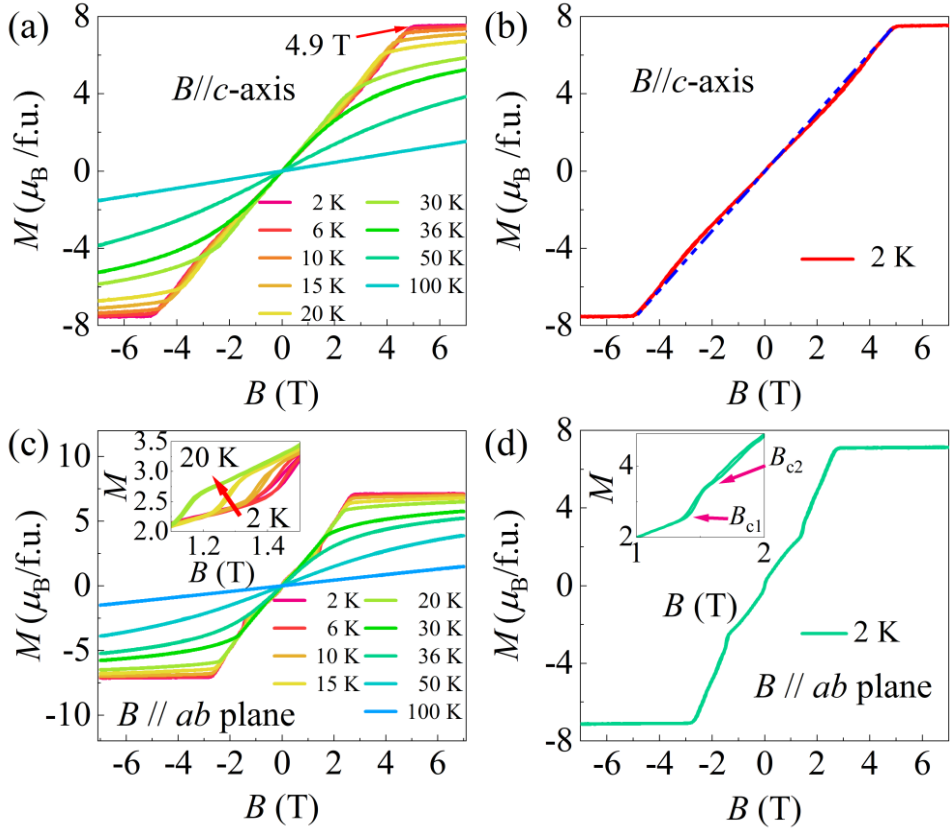

**Fig. S3.** (a) and (b) are isothermal magnetizations for EuAl$_2$Si$_2$ crystals with $B//c$-axis at different temperatures between 2 and 100 K. (c) and (d) are those with $B//ab$-plane at various temperatures between 2 and 100 K. Enlarged view around 1.2 T by the inset in (c) shows the temperature dependent spin flop and the one at 2 K is enlarged by the inset in (d).

The isothermal magnetizations of EuAl$_2$Si$_2$ at various temperatures between 2 and 100 K with the magnetic field along *ab*-plane and *c*-axis are shown in Figs. S4. The $B_{sat}$ are 4.9 T and 2.8 T at 2 K for *B*//*c*-axis and *B*//*ab*-plane, respectively, thus revealing the anisotropic magnetizations again. The saturation moments along both directions are rather close to spin only theoretical value (Eu$^{2+}$: ~7 $\mu_B$/f.u.). The data measured at 2 K are presented in Figs. S3(b) and S3(d) to show more details. When *B*//*c*-axis, the isothermal magnetizations slightly deviate from the linear evolution and show nearly invisible spin flops before spin polarization. While when *B*//*ab*-plane, two spin flop transitions are clearly visible near $B_{c1}$ = 1.4 T and $B_{c2}$ = 1.6 T.

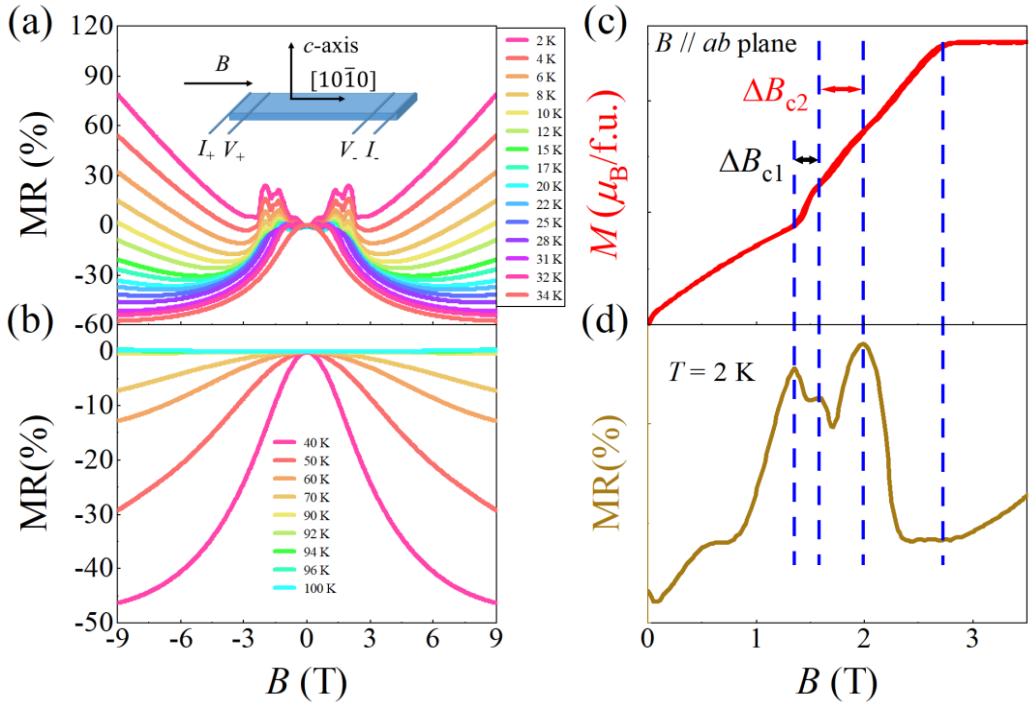

**Fig. S4.** (a) and (b) Magnetoresistance (*B*//*I* and *B*⊥*c*-axis) of EuAl$_2$Si$_2$ single crystals as a function of *B* at different temperatures. (c) Isothermal magnetizations of EuAl$_2$Si$_2$ measured at 2 K with *B*//*ab*-plane. (d) Magnetic field dependent magnetoresistance of EuAl$_2$Si$_2$ measured at 2 K with *B*//*ab*-plane.

The behavior of MR is closely related to the magnetizations. Seen in Fig. S4 by the magnetic field dependence of MR with *B*//*ab*-plane and *B*⊥*I* where *I* denotes the electrical current *I*, a crossover from negative to positive MR at $T_N$ is displayed.

Moreover, the MR exhibits multiple transitions caused by spin flops below 2.8 T. The inset of Fig. S4(a) shows the schematic configuration for MR measurements. The isothermal magnetizations and magnetic field dependence of MR at 2 K are presented in Figs. S4(c) and S4(d), respectively. It is clear that the kinks in the magnetic field dependent MR coincide well with the two spin flops seen in isothermal magnetizations, which could be guided by the dashed lines.

## 5. Quantum oscillation analysis

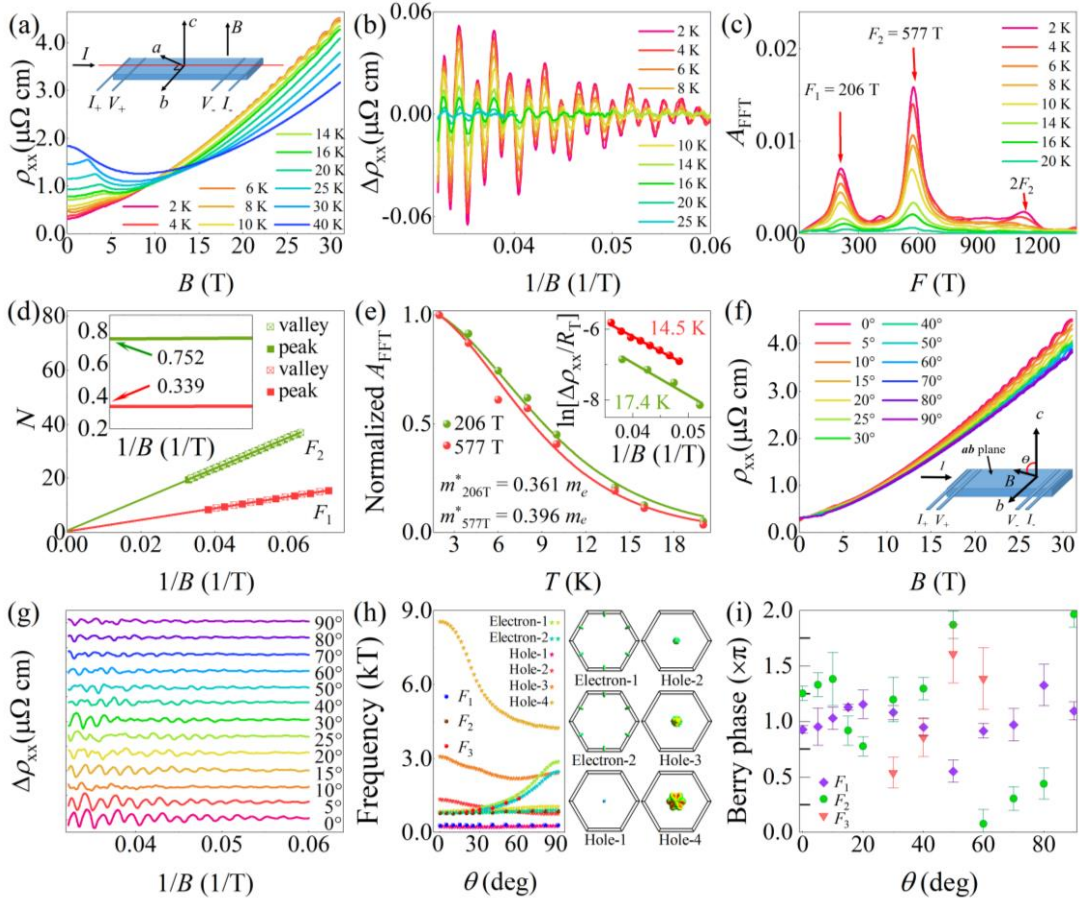

**Fig. S5.** (a) $\rho_{xx}$ vs. $B$ between 2 K and 40 K. Inset shows schematic measurement configuration. (b) The SdH oscillatory component vs. $1/B$ after subtracting the background. (c) FFT spectra of $\rho_{xx}$. (d) Landau index $N$ plotted against $1/B$ at 2 K. Inset enlarges intercepts of the fitting. Triangles represent the valley positions and squares denote the peak positions of SdH oscillations. (e) Temperature dependence of relative FFT magnitude of the two fundamental frequencies. The solid lines denote

the fitting results by using the L-K formula. Inset: Dingle plot of SdH oscillations at 2 K. (f) $\rho_{xx}$ vs. $B$ at various $\theta$ and 2 K. Inset shows schematic measurement configuration. (g) SdH oscillatory components vs. $1/B$ at various $\theta$ and 2 K. (h) Left: quantum oscillatory frequencies vs. $\theta$ between $B$ and $c$-axis at 2 K. $F_1$, $F_2$ and $F_3$ represent the experimentally derived fundamental frequencies, while pentagrams are the calculated values. The experimental frequency shown in the figure is multiplied by a factor of 1.35. Right: the calculated Fermi surface, including four hole-pockets and two electron-pockets. (i) The angle dependent Berry phase for $F_1$, $F_2$, and $F_3$. The error bar of the intercepts is determined from fitting.

To achieve more insights into the electronic band structure, magnetic field dependent $\rho_{xx}$ measured between 2 K and 40 K under high magnetic field up to 31 T is provided in Fig. S5(a). The MR, as shown in Fig. S6(a), reaches ~ $1.37 \times 10^3$% at 2 K and 31 T without showing any sign of saturation. Striking Shubnikov-de Haas (SdH) quantum oscillations in $\rho_{xx}$ are visible when $B \geq 15$ T. After carefully subtracted a fourth-power polynomial background, the SdH oscillations at different temperatures from 2 K to 25 K against the reciprocal magnetic field $1/B$ are presented in Fig. S5(b). The Quantum oscillations could be well described by the Lifshitz-Kosevich (L-K) equation [37]:

$$\Delta\rho_{xx} \propto R_S R_T R_D \cos\left[2\pi\left(\frac{F}{B} + \varphi\right)\right]$$

where $R_S = \cos(\pi g m^*/2m_e)$, $R_T = (2\pi^2 k_B T/\hbar\omega_c)/\sinh(2\pi^2 k_B T/\hbar\omega_c)$ and $R_D = \exp(-2\pi k_B T_D/\hbar\omega_c)$, representing the damping factors due to spin splitting, temperature, and scattering, respectively. The $k_B$ is the Boltzmann constant, $\hbar$ is the reduced Planck's constant, $F$ is the fundamental frequency of oscillation, $\varphi$ is the phase shift, $\omega_c = eB/m^*$ is the cyclotron frequency with $m*$ denoting the effective cyclotron mass, $T_D$ is the Dingle temperature defined by $T_D = \hbar/(2\pi k_B \tau_Q)$ with $\tau_Q$ being the quantum scattering lifetime. The fast Fourier transform (FFT) spectra of the SdH oscillations, depicted in Fig. S5(c), disclose two frequencies, $F_1 = 206$ T and $F_2 = 577$ T. This fact indicates that at least two pockets are across the Fermi level $E_F$. Fig. S6(b) shows the two filtered oscillatory parts of $\Delta\rho_{xx}$ at 2 K. The effective cyclotron mass $m^*$ at $E_F$ can be estimated from the L-K fitting to $R_T$, as is shown in Fig. S5(e), giving $m^* = 0.361\ m_e$ and $0.396\ m_e$ for $F_1$ and $F_2$, respectively, where $m_e$ denotes the free electron mass. The field dependent amplitudes of the quantum oscillations at 2 K are fitted by using the L-K formula, as shown in the inset to Fig. S5(e), giving the Dingle temperature $T_D = 17.4$ K and 14.5 K, respectively. The two fundamental frequencies $F_1$ and $F_2$ correspond to the external cross-sectional areas of the Fermi surfaces as $A = 1.96$ and 5.49 nm$^{-2}$, respectively, which are calculated by using the Onsager relation $F = (\hbar/2\pi)\ A$. The Fermi wave vectors for $F_1$ and $F_2$ are estimated to be 0.79 nm$^{-1}$ and 1.32 nm$^{-1}$, respectively, from $k_F = \sqrt{2eF/\hbar}$ and the corresponding Fermi velocities $v_F = 2.5 \times 10^5$ and $3.8 \times 10^5$ m s$^{-1}$ are calculated from $v_F = \hbar k_F/m^*$. The quantum mobility $\mu_Q \sim$ 340.9 and 372.96 cm$^2$ V$^{-1}$ s$^{-1}$ could also be obtained from $\mu_Q = e\tau_Q/m^*$. The corresponding quantum scattering lifetime $\tau_Q$ are $7.0\times10^{-14}$ and $8.4\times10^{-14}$ s, respectively. Other derived parameters are summarized in Table S3.

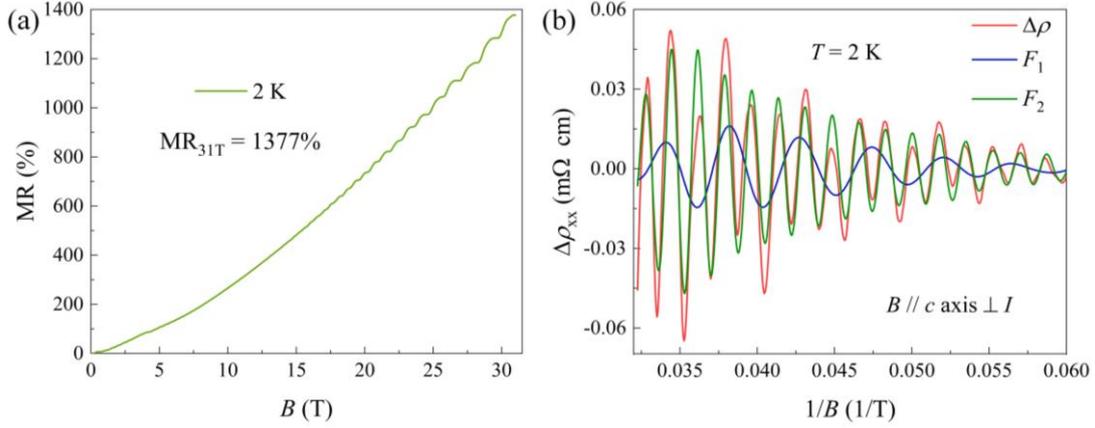

**Fig. S6.** (a) MR accompanied by distinct SdH oscillations at 2 K. (b) The two filtered oscillatory parts of $\Delta\rho_{xx}$. The red line represents raw SdH oscillatory signal of EuAl$_2$Si$_2$ with $B//c$-axis.

To understand the SdH oscillations, the Landau level (LL) index fan diagram is constructed, aiming to examine the Berry phase of EuAl$_2$Si$_2$ accumulated along the cyclotron orbit [38, 39]. In the L-K equation, $\gamma$ (= 1/2 - $\phi_B/2\pi$) is the Onsager phase factor and $\delta$ represents the FS dimension-dependent correction to the phase shift, which is 0 for 2D system and ± 1/8 (- for the electron-like pocket and + for the hole-like pocket) for 3D system [5, 39]. The LL index phase diagram is shown in Fig. S5(d), in which the valley positions of $\Delta\rho_{xx}$ against $1/B$ were assigned to be integer indices and the peak positions of $\Delta\rho_{xx}$ were assigned to be half-integer indices [40]. It is apparent that all the points nicely fall on a straight line, thus allowing a linear fitting that gives the intercepts of 0.339 and 0.752 corresponding to $F_1$ and $F_2$, respectively. The two intercepts correspond to the Berry phase $\phi_{B-1}$ of $2\pi$ (0.339+1/8) = 0.928$\pi$ and $\phi_{B-2}$ of $2\pi$ (0.752 - 1/8) = 1.254$\pi$, respectively. The Berry phase for $F_1$ is close to $\pi$, indicating the nontrivial topological nature of hole-1 band assigned by the first-principles calculations presented later. As we noted above, during the $\rho_{xx}$ measurements, $B$ was initially perpendicular to the $ab$-plane, i.e. parallel to the $c$-axis. To map out the whole Fermi surface, we rotated the $c$-axis of the crystal gradually deviating from $B$. The rotation geometry is depicted in the inset of Fig. S5(f), in which $\theta$ is the angle between $B$ and the $c$-axis. The angle dependent $\rho_{xx}$ is presented in Fig. S5(f). Fig. S5(g) present the angle dependent SdH oscillations with a constant offset in the vertical coordinate, which display clear shift of the peaks and the peak numbers are obviously changed with the increases of $\theta$. When the magnetic field was rotated from $B//c$-axis gradually to $B//b$-axis, the experimentally determined $F_1$ and $F_2$ were compared with theoretical values, as shown in the left inset to Fig. S5(h), changing from 206 T and 577 T at $\theta$ = 0° (out of plane) to 209 T and 628 T at $\theta$ = 90° (in plane). The experimental frequencies in Fig. S5(h) are multiplied by a factor of 1.35 to guide the eyes. However, when $\theta$ is between 30° and 60°, a new frequency $F_3$ appears, thus unambiguously unveiling the anisotropy of the Fermi pockets associated with the SdH oscillations. In the theoretical calculations, the Fermi surface based on the mBJ band structure of $c$-axis spin polarized FM-c EuAl$_2$Si$_2$ (see Fig. S12) consists of two electron pockets and four hole pockets, as shown in the right inset to Fig. S5(h). By a careful comparison between theoretical calculations and experimental results, we could reach a conclusion that the $F_1$ comes from the hole-1 pocket, which forms the Weyl point (WP), while $F_2$ and $F_3$ come from the electron-2 pocket.

Table S3. Parameters derived from SdH oscillations for EuAl$_2$Si$_2$, where $\kappa_F$ is the Fermi wave vector, $v_F$ denotes the Fermi velocity, $\tau_Q$ is the relaxation time and $\varphi_B$ is the Berry phase.

|  | $F$ (T) | $A$ (nm$^{-2}$) | $K_F$ (nm$^{-1}$) | $v_F$ (m/s) | $m^*/m_e$ | $T_D$ (K) | $\tau_Q$ (s) | Berry phase |
|---|---|---|---|---|---|---|---|---|
| $F_1$ | 206 | 1.96 | 0.79 | 2.5×10$^5$ | 0.361 | 17.4 | 7.0×10$^{-14}$ | 0.928$\pi$(+1/8) |
| $F_2$ | 577 | 5.49 | 1.32 | 3.8×10$^5$ | 0.396 | 14.5 | 14.5×10$^{-14}$ | 1.254$\pi$(-1/8) |

To further reveal the band structure topology of EuAl$_2$Si$_2$, we present the Berry phase at various tilt angles, as shown in Fig. S5(i). The Berry phase of $F_1$ band shows strong anisotropy, which at the tilt angles between 0° to 40°, 60°, 70° and 90° fluctuates between 0.9$\pi$ and 1.15$\pi$, suggestive of nontrivial band structure topology. However, when $\theta$ is between 50° and 80°, the Berry phase are 0.55$\pi$ and 1.32$\pi$, implying trivial band structure topology. For $F_2$ and $F_3$ bands, the Berry phases are apparently away from $\pi$, indicating trivial band structure topology.

## 6. First-principles calculations

Electronic band structures for different spin configurations of EuAl$_2$Si$_2$ were calculated within the framework of the projector augmented wave (PAW) method [41] and employed either the generalized gradient approximation (GGA) with Perdew-Burke-Ernzerhof (PBE) [42] formula or the modified Becke-Johnson method [43] as implemented in the Vienna Ab initio Simulation Package (VASP) [44]. The experimental crystal structures were chosen for the whole calculations scheme. The Brillouin zone was sampled with 11×11×5 and 11×11×7 Monkhorst-Pack k-mesh for A-type AFM and FM structures, respectively. Kinetic energy cutoff was set to 400 eV and spin-orbital coupling (SOC) was self-consistently incorporated. In order to describe correlated interaction of Eu 4$f$ electrons, the Hubbard values were tested from 3 eV to

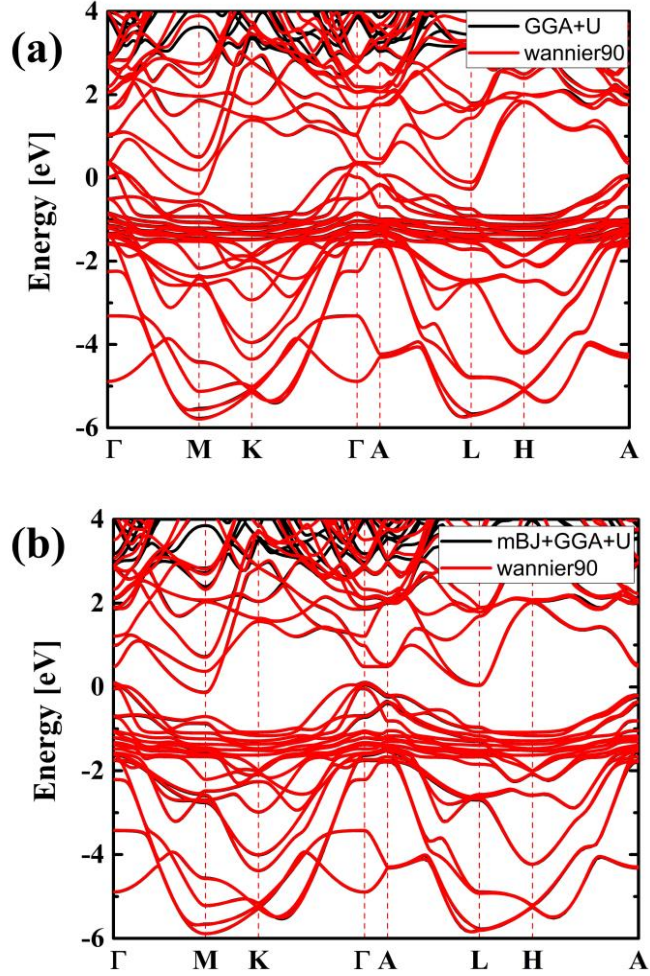

**Fig. S7.** The wannier90 projected band structures of in-plane A-type AFM EuAl$_2$Si$_2$ with (a) GGA + U = 5 eV and (a) mBJ + GGA + U = 5 eV.

8 eV. Magnetic space groups of EuAl$_2$Si$_2$ at the specific magnetic orders were determined by using FindSym [45, 46], BilBao crystallographic server [47], and output by VASP. Magnetic topological quantum chemistry method was used to classify the topological properties for EuAl$_2$Si$_2$ by using MagVasp2trace package [48-50]. Eu 4$f$5$d$, Al 3$s$3$p$, and Si 3$p$ orbitals are projected onto maximally localized Wannier functions using VASP2WANNIER interface [51], as shown in Figs. S7 and S8, and Berry curvature was calculated by postw90.x [52]. Corresponding topological properties were calculated by using WANNIERTOOLS [53] and bulk Fermi surfaces were plotted by FermiSurfer [54] software. The intrinsic anomalous Hall conductivity is determined by the band geometric properties. We calculate the intrinsic anomalous Hall conductivity defined by

$$\sigma_{xy}^A = -\frac{e^2}{\hbar}\sum_n \int \frac{d\mathbf{k}}{(2\pi)^3} f_n(\mathbf{k})\Omega_{n,z}(\mathbf{k}).$$

Here $\Omega_{n,z}(\mathbf{k})$ is z component of Berry curvature

$$\Omega_{n,z}(\mathbf{k}) = -2\mathrm{Im}\sum_{n\neq n'}\frac{<n\mathbf{k}|v_x|n'\mathbf{k}><n'\mathbf{k}|v_y|n\mathbf{k}>}{(\omega_{n'}-\omega_n)^2}$$

where $n$ ($n'$) is band index and $v$ refers to velocity operator. And the band energy of $n$-th band is $\hbar\omega_n$. We calculated the quantum oscillation frequencies via adopting the SKEAF package [55].

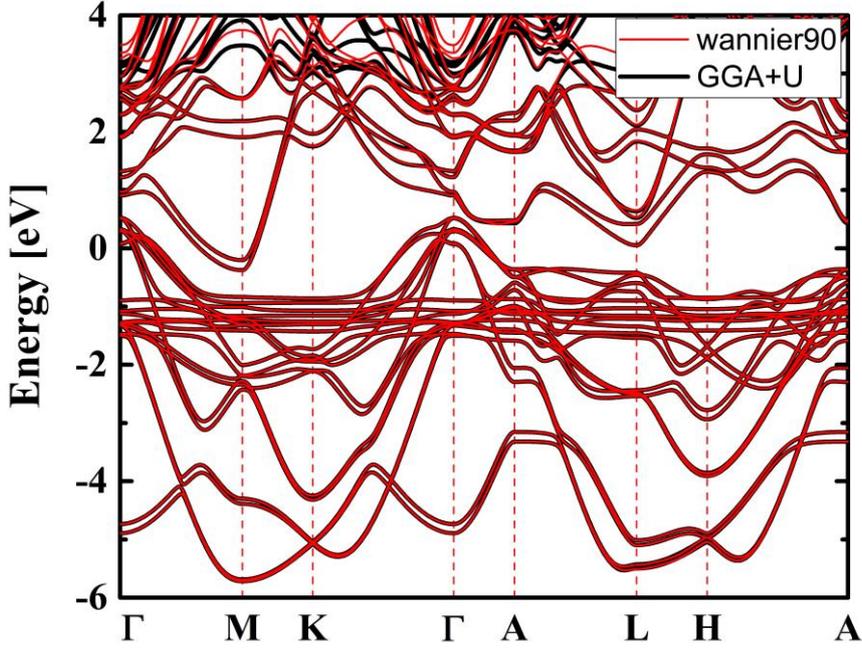

**Fig. S8.** The wannier90 projected band structures of FM-c type EuAl$_2$Si$_2$ with (a) GGA + U = 5 eV.

**Table S4.** Parity eigenvalues at eight inversion-invariant momenta. The format is positive/negative parity results ($n_+/n_-$). Total electron number is 62 for in-plane A-type AFM and 31 for FM-out.

|  | (0,0,0) | (0.5,0,0) | (0,0.5,0) | (0.5,0.5,0) | (0,0,0.5) | (0.5,0,0.5) | (0,0.5,0.5) | (0.5,0.5,0.5) | $Z_4$ |
|---|---|---|---|---|---|---|---|---|---|
| AFM-62 | 22/40 | 20/42 | 20/42 | 20/42 | 31/31 | 31/31 | 31/31 | 31/31 | 2 |
| FM-31 | 10/21 | 10/21 | 10/21 | 10/21 | 21/10 | 21/10 | 21/10 | 21/10 | 0 |
| FM-30 | 10/20 | 10/20 | 10/20 | 10/20 | 21/9 | 20/10 | 20/10 | 20/10 | 1 |
| FM-29 | 10/19 | 10/19 | 10/19 | 10/19 | 21/8 | 19/10 | 19/10 | 19/10 | 2 |
| FM-28 | 10/18 | 10/18 | 10/18 | 10/18 | 21/7 | 18/10 | 18/10 | 18/10 | 3 |

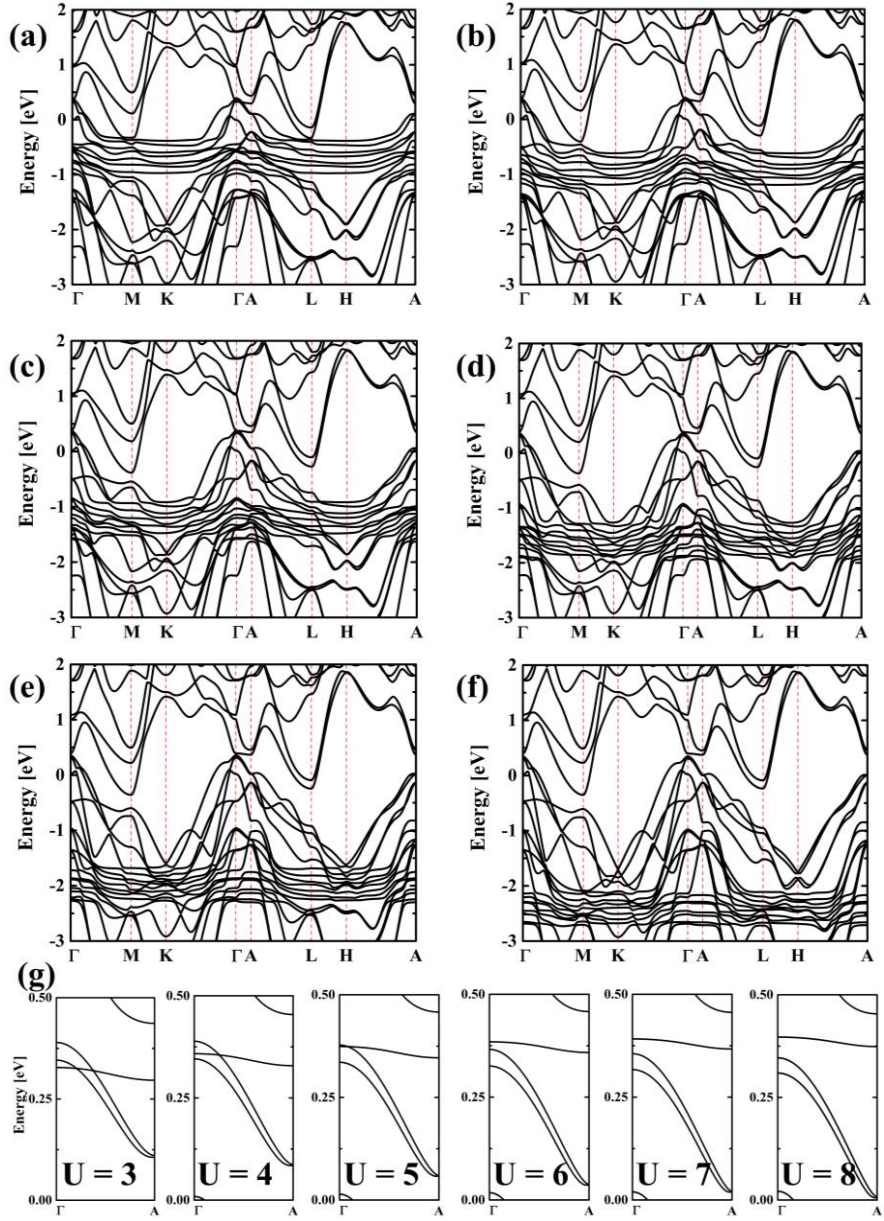

**Fig. S9.** The band structures of in-plane A-type AFM EuAl$_2$Si$_2$ by using GGA + U method with (a) U = 3 eV; (b) U = 4 eV; (c U = 5 eV; (d) U = 6 eV; (e) U = 7 eV; (f) U = 8 eV; (g) enlarge region along high symmetry line Γ-A of the Brillouin zone.

The ground state magnetic structure was chosen by using the first-principles calculations. First, we constructed ferromagnetic (FM) and A, C, G-type antiferromagnetic (AFM) spin configurations of EuAl$_2$Si$_2$ with spins along out-of-plane and in-plane directions. Based on total energy calculations, it is found that the magnetic ground state is an in-plane A-type AFM one, which is consistent with our measured isothermal magnetizations (See Fig. S3(a,c)) as well as with previous neutron diffraction measurements results [56]. Second, we also compared three spin directions of in-plane A-type AFM along the *a*-, *b*-, and [110] axes, in which total energy difference can be less than $10^{-5}$ eV. This result may support complex spin flop transition within the *ab*-plane (see Fig. S3(d)). In the following calculations, the spin direction of Eu is set as along the *b*-axis for consistency. According to symmetry analysis, the magnetic space group of this A-type AFM EuAl$_2$Si$_2$ belongs to type-IV $C_c2/c$ with BNS setting (Belov-Neronova-Smirnova notation), which keeps inversion symmetry but breaks $C_{3z}$ and time-reversal symmetries. Given these symmetry operators, it can still hold Z$_4$ topological numbers based on parity values [57, 58], which can be calculated based on formula:

$$Z_4 = \sum_{k=1}^{8} \sum_{n=1}^{n=n_{occ}} (1 + p_k^n)/2 \mod 4,$$

where $p_k^n$ are the parity eigenvalue (+1/-1) for occupied band *n* at eight inversion-invariant momenta *k*. Z$_4$ = 1 or 3 means Weyl semimetal state and Z$_4$ = 2 is axion insulator. The parity eigenvalues of $C_c2/c$ were calculated based on magnetic topological quantum chemistry (MTQC) [48-50] with GGA+U (U = 5 eV), with the results shown in Table. S4. The corresponding trace file of this magnetic space group was calculated, which gives the result of a nontrivial topological insulator [59]. Both calculations results present that Z$_4$ topological index is 2, which belongs to axion insulator, resembling the case of A-type AFM EuCd$_2$As$_2$ [60]. However, the gap due to $C_{3z}$ symmetry breaking is very small and energy position is higher than the E$_F$, which hinders the detection of the axion insulator state. Because of correlation effect of Eu orbital, we calculated the band structures with different Hubbard U values from 3 to 8 eV, as illustrated in Fig. S9. Based on MTQC analysis, axion insulator only appears at U equals to 4 and 5 eV, where U values larger than 5 eV destroy the band inversion. Meantime, the mBJ potential also increases the gap without band inversion (Fig. S10).

Therefore, the topological properties of in-plane A-type AFM EuAl$_2$Si$_2$ show Hubbard U sensitive characteristics.

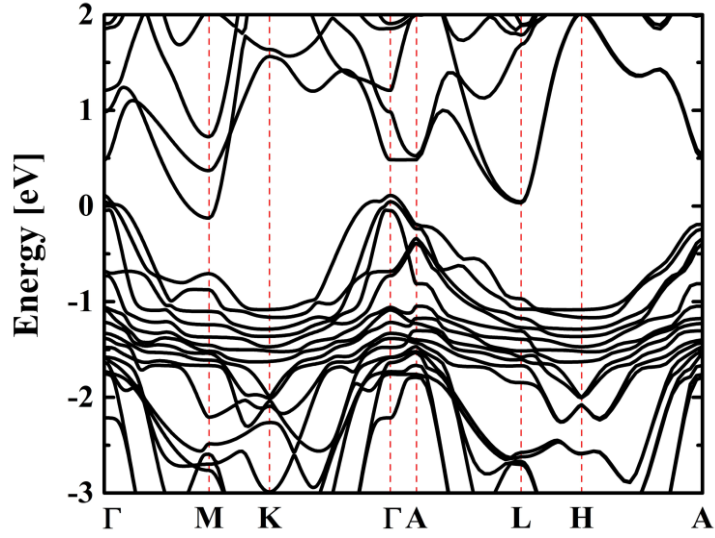

**Fig. S10.** The mBJ band structure of in-plane A-type AFM EuAl$_2$Si$_2$ with GGA + U = 5 eV.

The applied strong magnetic field can polarize all Eu spins along the *c*-axis, which owns type-III $P\bar{3}m'1$ magnetic space group. For such spin polarized FM-c state, time-reversal symmetry is broken but spatial inverse symmetry is preserved. It is found that there is a gap between valance band and conduction band of the electronic band structure, indicating trivial insulator. Although it is trivial insulator for total occupied states (total valence electrons number $n = 31$), compatibility relationship supports symmetry protected linear band crossing points along Γ-A direction of the Brillouin zone for several top valence bands ($n = 28, 29, 30$), as shown in Table. S4. Based on these $Z_4$ results, there are WPs for these top valence occupied states as indicated in Table. S4. Meanwhile, these linear band crossing points are robust against Hubbard U values from 4 to 8 eV, as shown in Fig. S11. This is also examined by the mBJ method, showing that a pair of WPs is close to E$_F$ of about 57 meV, as shown in Fig. S12.

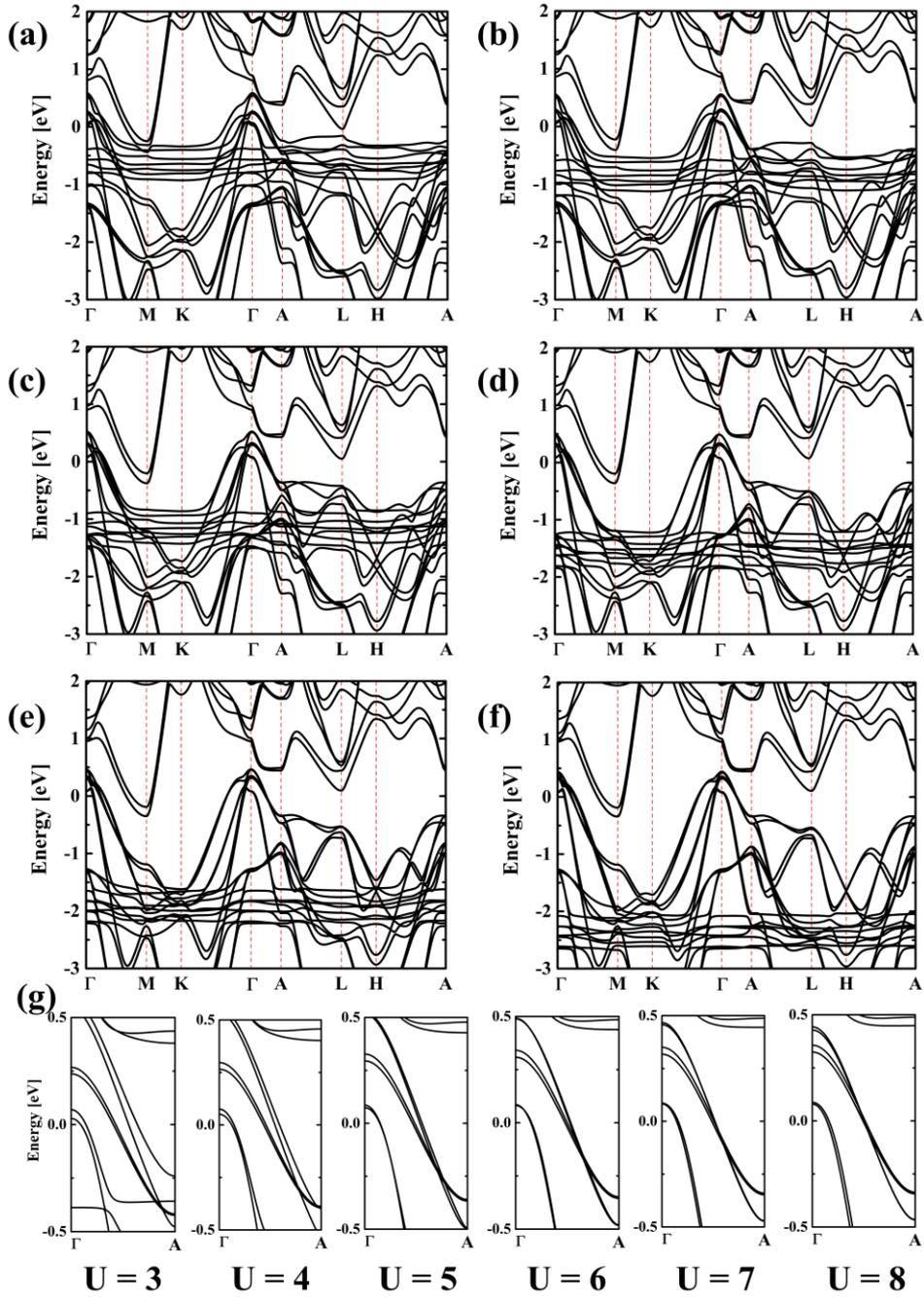

Fig. S11. The band structures of FM-c type EuAl$_2$Si$_2$ by using GGA + U method with (a) U = 3 eV; (b) U = 4 eV; (c) U = 5 eV; (d) U = 6 eV; (e) U = 7 eV; (f) U = 8 eV; (g) enlarge region along high symmetry line Γ- A.

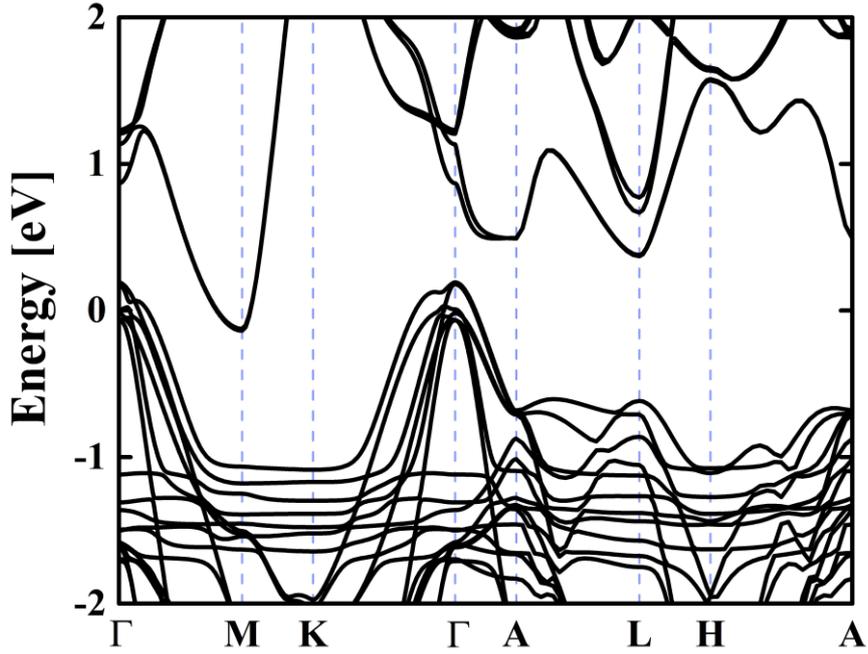

**Fig. S12.** The mBJ band structure of EuAl$_2$Si$_2$ with FM-c configuration.

## 7. Angle-resolved photoemission spectroscopy experiments

Angle-resolved photoemission spectroscopy (ARPES) measurements were performed at 03U beamlines of Shanghai Synchrotron Radiation Facility (SSRF) [61]. The samples were cleaved *in situ* and measured under ultrahigh vacuum below $8\times10^{-11}$ Torr. Data were collected by Scienta DA30 analyzer. The angular and the energy resolutions were set to 0.2° and 12～20 meV (dependent on the selected probing photon energy).

### 7.1 Electronic structure of in-plane A-type AFM EuAl$_2$Si$_2$

To directly detect the band dispersion near E$_F$, we performed high-resolution ARPES measurements on (001) cleavage surface of EuAl$_2$Si$_2$. Fig. S13(a) displays the angle-integrated photoemission spectrum, in which we can identify the Si-2*p*, Al-2*p* and Eu-4*f* peaks clearly. The sharp spectral feature confirms the high crystal quality and clean sample surface. The Eu-4*f* state has double-peak structure and the energy separation is about 0.9 eV, as shown by the valence band spectrum in Fig. S14. Then we carried out the photon-energy-dependent ARPES measurements to investigate the detailed band dispersions along *k$_z$* and determine the high symmetry surface. As shown in Fig. S13(b), a periodic modulation along the $k_z$ direction can be found. Thus, we can determine that $hv$ = 60, 91 and 130 eV are close to the Γ point, and 75 and 110 eV are close to the A point according to the free electron final-state model. Fig. S13(c) shows the Fermi surface mapping with the incident photon of 91 eV and the calculated results. The Fermi surface consists two hole-like pockets surrounding Γ point (labeled as α and β) and small electron-like pockets at M point (labeled as δ), which contributes to the typical semimetallic nature of this compound, consistent with our transport measurements.

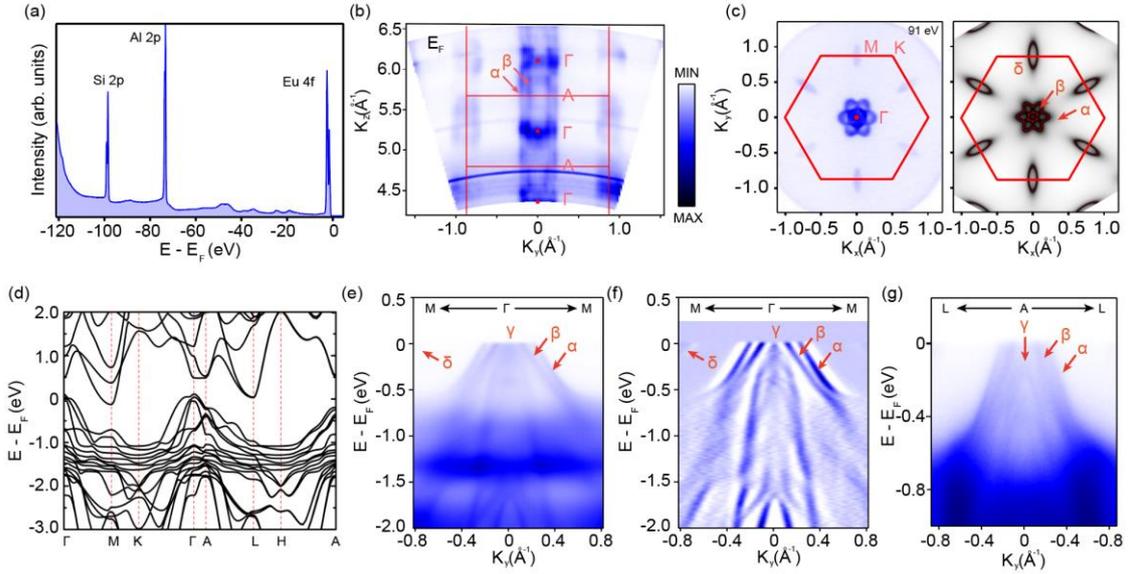

**Fig. S13.** The electronic structure of in-plane A-type AFM EuAl$_2$Si$_2$. (a) Core-level photoemission spectrum shows strong Si 2p, Al 2p and Eu 4f peaks. (b) Intensity plot of $k_z$ dependent ARPES data along Γ-M direction. (c) Fermi surfaces of EuAl$_2$Si$_2$: (i) Intensity plot at E$_F$ ± 20 meV taken with 91 eV photons (corresponding to $k_z$-π plane) and (ii) calculated Fermi surfaces. (d) The calculated electronic band structure. (e) and (f) Intensity plots and corresponding second-derivative plots along M-Γ-M direction taken with 91 eV photons. (g) Intensity plot along L-A-L direction taken with 110 eV photons.

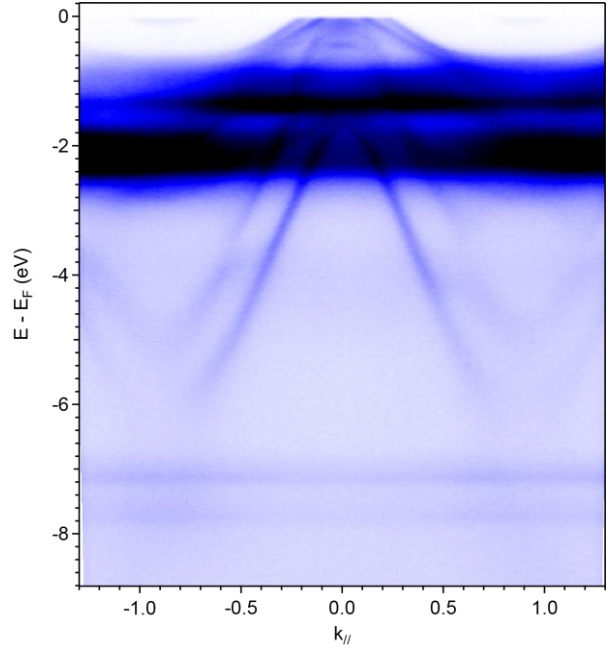

**Fig. S14.** Dispersions of valence band of EuAl$_2$Si$_2$ with incident photon energy of 91 eV.

The band structure near E$_F$ is dominated by Eu-5$d$, Al-3$s$ and Si-3$p$ orbits according to our atomic orbital-projected calculation. Fig. S13(d) declares the calculated band structure of in-plane A-type AFM EuAl$_2$Si$_2$ with mBJ potential. The results reveal a band gap at the center of Brillouin zone and the size is depended on the calculation method. Generalized gradient approximation (GGA+ U, U = 5 eV) calculations show a smaller nontrivial band inversion gap, which is defined by Z$_4$ topological invariant and is analogy to the case of A-type AFM EuCd$_2$As$_2$ [60]. Figs. S13(e) and (f) show the band dispersion measured at 12 K along the high-symmetry M-Γ-M direction and the corresponding second-derivative plot. One prominent feature is the flat bands located at -1 to -1.5 eV and -1.8 to -2.5 eV below E$_F$, which can be attributed to Eu 4$f$ orbitals. In the vicinity of E$_F$, we can identify three hole-like bands and one electron-like band. The outer two bands, designated as $\alpha$ and $\beta$, disperse linearly away from E$_F$ and appears as a sharp spectral feature at all incident photon energies, and the Fermi surfaces of these bands are just vertical lines along the $k_z$ direction. Therefore, we can conclude that $\alpha$ and $\beta$ bands are two-dimensional surface states. This feature is similar to that of other Eu-based axion insulator candidates, such as EuSn$_2$P$_2$, EuSn$_2$As$_2$ and EuIn$_2$P$_2$ [62-65]. In Fig. S13 (g), we show the photoemission intensity plots along the L-A-L high-symmetry path. Obviously, it is found that the top of $\gamma$ band shifts downward at the A point.

## 8. Magnetic force microscopy measurement

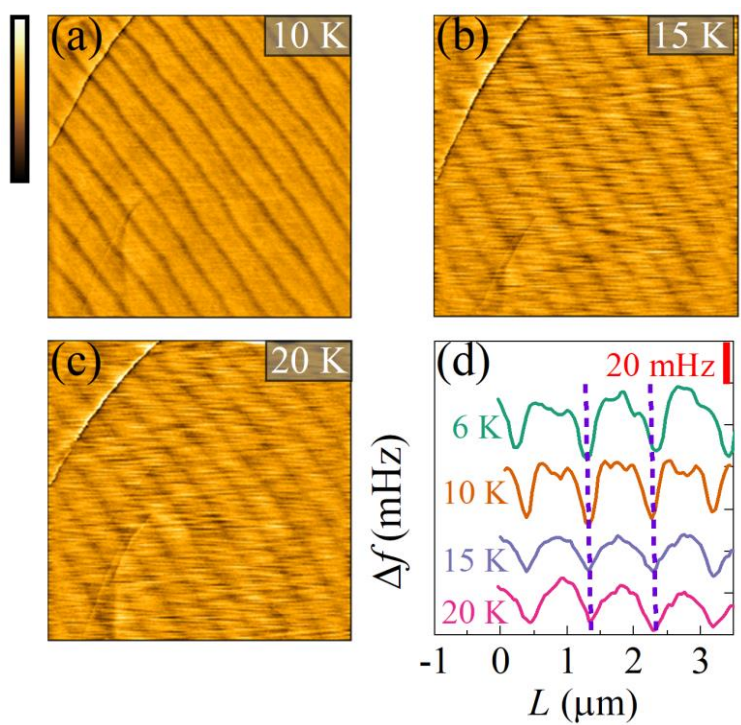

**Fig. S15.** (a)-(c) Zero-field MFM images of EuAl$_2$Si$_2$ at 10 K, 15 K and 20 K. The color scales in images (a)-(c) are 140 mHz. (d) A line profile of DW at 6 K, 10 K, 15 K and 20 K, where vertical purple lines mark the position of the DWs.

The EuAl$_2$Si$_2$ single crystal was cleaved in air to expose the (001) surface. After the cleavage, we used both optical microscope and atomic force microscopy to check the cleaved surface at room temperature in air to ensure its high quality. All the MFM pictures were taken using a cryogenic magnetic force microscopy under ambient pressure at different temperatures and magnetic fields. The MFM images were taken at a scale of 10 μm × 10 μm. 2 nm Ti and 10 nm Au film were deposited on the sample surface to eliminate electrostatic stray fields.

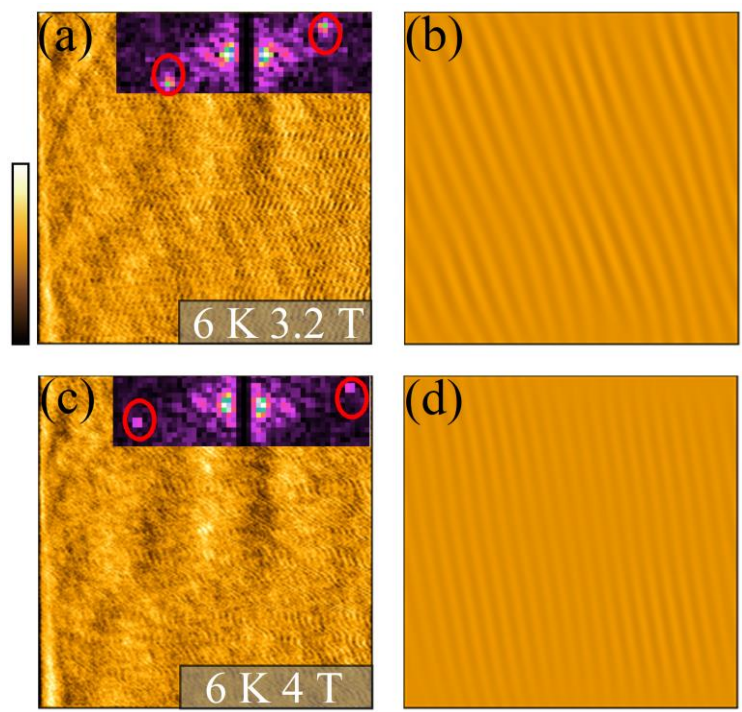

**Fig. S16.** (a) and (c) Striped magnetic configuration of EuAl$_2$Si$_2$ observed using MFM at 3.2 T and 4 T, respectively. Inset: Fast Fourier transform (FFT) pattern of the image. (b) and (d) Reconstructed MFM image from the inverse FFT after applying masks on the two reflections circled by red open circles in (a) and (c), respectively. The color scales in images (a)-(h) are 40 mHz.

Fig. S15 presents the MFM images of EuAl$_2$Si$_2$ taken at 10 K, 15 K and 20 K, which show clear stripe domain wall (DW) structure. The periodicity of the DWs is almost invariant with the change of temperature. Figs. S16(a) and 16(c) plot the MFM images at 6 K with different magnetic fields. In high fields, the stripe domain structure gradually becomes blurred, so we extract the period of the DW by FFT, as shown in the insets of Figs. S16(a) and 16(c), respectively. The MFM image was reconstructed from the inverse FFT after applying masks on the two reflections circled by red open circles in (a) and (c), respectively, as shown in Fig. S16(b) and 16(d). Fig. S17 shows the schematic view of skew scattering of Bloch waves at the DW of the two layers of EuAl$_2$Si$_2$ at 0 T.

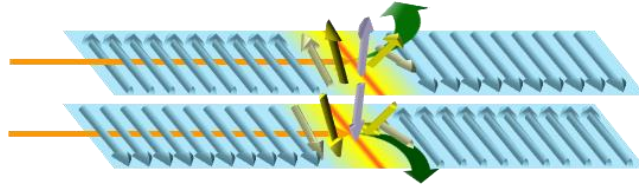

**Fig. S17** Schematic view of skew scattering of Bloch waves at the DW of the two layers of EuAl$_2$Si$_2$ at 0 T.

## 9. Analysis of the magnetotransport data

We then turn to analyze the magnetotransport data. To eliminate the influence of $\rho_{xx}$ on $\rho_{xy}$, $\rho_{xy}$ is symmetrized by using $\rho_{xy} = [\rho_{xy}(+B) - \rho_{xy}(-B)]/2$. If it exhibits a linear dependence on $B$ when $B > B_N$, the electron or hole carriers dominate the transport, thus resulting in a constant $R_H$ [5, 6, 66-69]. While when $B > B_N$, if $\rho_{xy}$ is no longer linear, the fitting should employ the standard two-band model [70, 71] expressed as

$$R_H(B) = \frac{1}{e} \cdot \frac{(n_h\mu_h^2 - n_e\mu_e^2) + (n_h - n_e)\mu_e^2\mu_h^2 B^2}{(n_h\mu_h + n_e\mu_e)^2 + (n_h - n_e)^2\mu_e^2\mu_h^2 B^2},$$

where $R_H$ is the Hall coefficient ($R_H = \rho_H/B$), $n_e(n_h)$ denotes the carrier density for the electron (hole), and $\mu_e(\mu_h)$ is the mobility of the electron (hole). In our case, the two-band model was employed for the fitting. The Hall conductivity can be calculated by using $\sigma_{xx} = \rho_{xx}/(\rho^2_{xy} + \rho^2_{xx})$ and $\sigma_{xy} = -\rho_{xy}/(\rho^2_{xy} + \rho^2_{xx})$ and $\sigma^{DW}_{xy} = -\rho^{DW}_{xy}/(\rho^2_{xy} + \rho^2_{xx})$, respectively. Figs. S18 and S19 show the fitting results indicated by the solid lines between 2 and 25 K and between 34 and 300 K, respectively.

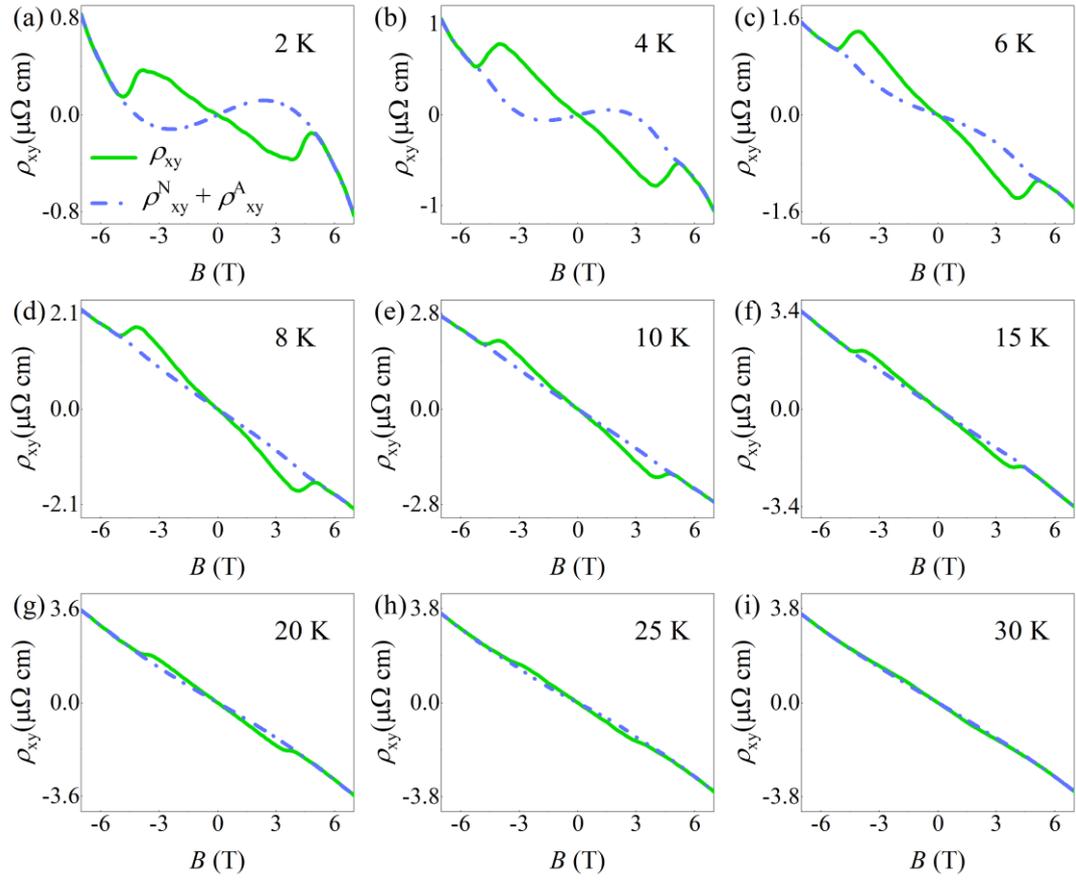

**Fig. S18.** (a)-(i) Hall resistivity (green lines) fitted by using $\rho_{xy} = \rho^N_{xy} + \rho^A_{xy}$ (blue solid lines) between 2 and 25 K and $B//c$ and $B \perp I$. The clear deviation from conventional Hall resistivity indicates the presence of an additional term.

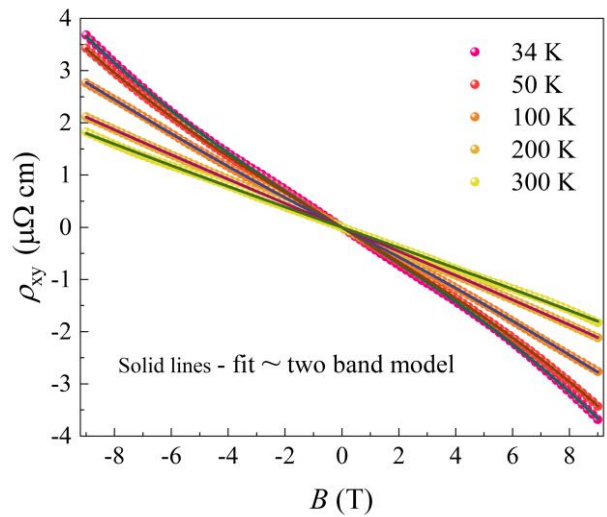

**Fig. S19.** Analysis of $\rho_{xy}$ by using the two-band model. (a) Hall resistivity *vs*. B between 34 and 300 K. The solid lines denote the fitting results by using the two-band model.

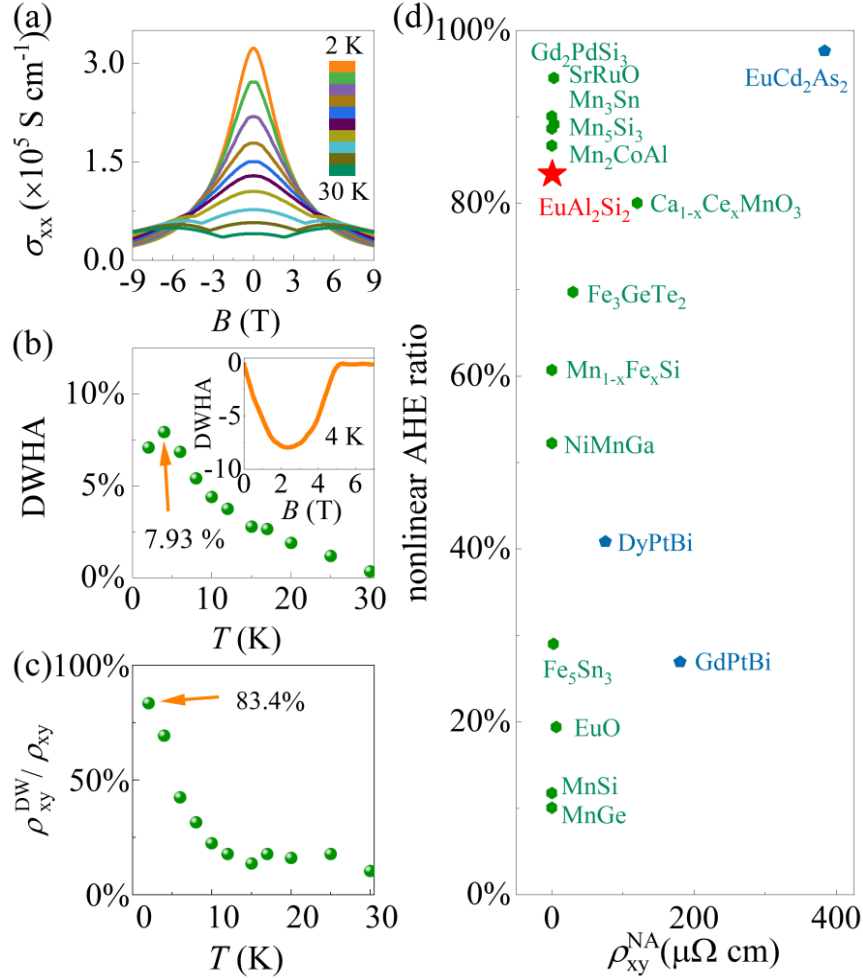

**Fig. S20.** A comparison of the anomalous Hall conductivity of EuAl$_2$Si$_2$ with other AHE systems. (a) Magnetic field dependent conductivity $\sigma_{xx}$ at different temperatures. (b) The maximum DWHA $\theta^{DW}_{xy}$ of EuAl$_2$Si$_2$ at different temperatures. Inset: Magnetic field dependent $\theta^{DW}_{xy}$ at 4 K. (c) The maximal DWHE ratio $\rho^{DW}_{xy}/\rho_{xy} \times 100\%$ at different temperatures. (d) Nonlinear AHE ratio of intrinsic AHE systems (blue) and topological Hall effect systems (green). The DWHE in EuAl$_2$Si$_2$ contributes 83.4% of the total Hall signal at 2 K.

**Table S5** A summary of experimentally identified anomalous Hall systems.

| | System | $\sigma_{xx}$ (S cm$^{-1}$) | $\sigma_{AH}$ (S cm$^{-1}$) | AHA% ($\sigma_{AH}/\sigma$) | Ref |
|---|---|---|---|---|---|
| FM | L1$_0$-FePt(f) | 3.79×10$^4$ | 1250 | 3.3 | 72 |
| | MnGa(f) | 5100 | 288 | 5.7 | 73 |
| | TbCo(f) | 2.5×10$^4$ | 800 | 3.2 | 74 |
| | SmFe(f) | 6600 | 317 | 4.8 | 75 |
| | Ga$_{1-x}$Mn$_x$As(f) | 2200 | 147 | 6.6 | 76 |
| | Fe (f) | 4.3×10$^4$ | 1134 | 2.6 | 23 |
| | Gd(f) | 5×10$^5$ | 1000 | 0.2 | 23 |
| | Cu$_{1-x}$Zn$_x$Cr$_2$Se$_4$ | 5×10$^4$ | 500 | 1 | 23 |
| | Mn$_5$Ge$_3$ | 1.41×10$^4$ | 860 | 6.1 | 77 |
| | Co$_3$Sn$_2$S$_2$ | 5700 | 1130 | 19.8 | 78 |
| | Co$_2$MnGa | / | 1530 | / | 79 |
| | Co$_2$MnAl | / | 2000 | / | 80 |
| | MnSi | 4000 | 150 | 3.8 | 8 |
| | ZrMnP | / | 2000 | 10.2 | 81 |
| | HfMnP | / | 2840 | 13.6 | 81 |
| | Fe$_3$Sn$_2$ | 105 | 1100 | 1.1 | 82 |
| | LaCrSb$_3$ | / | 1250 | 4 | 83 |
| | NdCrSb$_3$ | / | 2900 | 7.7 | 83 |
| | CeCrSb$_3$ | / | 1550 | 10 | 83 |
| | TbMn$_6$Sn$_6$ | / | 230 | 2.3 | 84 |
| | DyMn$_6$Sn$_6$ | / | 250.3 | 2.2 | 84 |
| | CrTe$_2$ (flake) | 1.3×10$^6$ | 67000 | 5.5 | 85 |
| AFM | Mn$_3$Sn | 3100 | 100 | 3.2 | 9 |
| | Mn$_3$Ge | 9000 | 450 | 5 | 11 |
| | Mn$_3$Pt | / | 98 | / | 12 |
| | RuO$_2$ | / | 330 | / | 86 |
| | Mn$_5$Si$_3$ | / | 102 | / | 87 |
| | CoNb$_3$S$_6$ | / | 400 | / | 88 |
| | GdPtBi | 700 | 110 | 16 | 17 |
| | NdPtBi | / | 110 | 16 | 17 |
| | TbPtBi | / | 744 | 38 | 17 |
| | MnBi$_2$Te$_4$ | / | 2.5 | / | 89 |
| | EuCd$_2$As$_2$ | 6250 | 100 | 1.6 | 90 |
| | EuCd$_2$Sb$_2$ (f) | 454 | 51.3 | 11.3 | 91, 15 |
| | ErMn$_6$Sn$_6$ | / | 80-308 | 1.7 | 84 |
| | EuAl$_2$Si$_2$ | 3.22×10$^5$ | $\sigma_{DW}$: 1.51×10$^4$ | $\sigma_{DW}/\sigma$: 7.93 | This work |

$\sigma_{xx}$: longitudinal charge conductivity; $\sigma_{AH}$: anomalous Hall conductivity; $\sigma_{DW}$: DW Hall conductivity; **AHA:** anomalous Hall angle; **f** denotes film. Some AHA values are calculated by using the data from the references. Here we use the absolute amplitudes of $\sigma_{AH}$ without the negative sign. Only the maximum values of AHA are adopted here.

A comparison between our results of $\sigma^{DW}_{xy}$ dependent DW Hall angle (DWHA) and previously reported $\sigma^A_{xy}$ dependent AHA for other AHE materials was presented in Table S5 [8, 9, 11, 12, 23, 72-91]. The AHC values of various films and thin flakes are also summarized. The magnetic field dependent Hall conductivity $\sigma_{xx}$ at different temperatures is shown in Fig. S20(a), which unveils that EuAl$_2$Si$_2$ has a very large $\sigma_{xx}$ of ~ $8.33 \times 10^5$ S cm$^{-1}$ at 2 K and 0 T. Moreover, DWHA $\theta^{DW}_{xy}$ ($\theta^{DW}_{xy} = \sigma^{DW}_{xy}/\sigma \times 100\%$) shown in Fig. S20(b) at 4 K reaches the maximum value of 7.93%. The DWHE ratio, defined as $\frac{|\rho^{DW}_{xy}|}{|\rho^{N}_{xy}|+|\rho^{A}_{xy}|+|\rho^{DW}_{xy}|}$ [40], reaches ~ 83.4% at 2 K as shown in Fig. S20(c), indicating that most of the Hall effect is contributed by the DWHE at the peak position. In Fig. S20(c) and Table S6, the nonlinear anomalous Hall effect (AHE) ratio is compared with previous reported values of other AHE systems [17, 32, 67, 92-104].

Table S6 Nonlinear AHE ratio of intrinsic AHE systems.

| System | $\rho^{NA}_{xy}$ (μΩ cm) | nonlinear AHE ratio % | Ref |
|---|---|---|---|
| MnSi | 0.004 | 11.7 | 32 |
| MnGe | 0.16 | 10 | 93 |
| EuO | 6 | 19.4 | 94 |
| Fe$_5$Sn$_3$ | 2.1 | 29 | 95 |
| GdPtBi | 180 | 26.9 | 17 |
| DyPtBi | 75 | 40.8 | 96 |
| NiMnGa | 0.14 | 52 | 67 |
| Mn$_{1-x}$Fe$_x$Si | 0.03 | 60.7 | 97 |
| Fe$_3$GeTe$_2$ | 30 | 69.7 | 98 |
| Ca$_{1-x}$Ce$_x$MnO$_3$ | 120 | 80 | 99 |
| Mn$_2$CoAl | 0.007 | 86.7 | 100 |
| Mn$_5$Si$_3$ | 0.003 | 88.7 | 101 |
| Mn$_3$Sn | 3.1 | 89 | 102 |
| SrRuO$_3$ | 0.2 | 90 | 103 |
| Gd$_2$PtSi$_3$ | 2.6 | 94.5 | 104 |
| EuCd$_2$As$_2$ | 383.48 | 97.6 | 92 |
| EuAl$_2$Si$_2$ | 0.49 | 83.4 | This work |

The magnetic field dependent Hall resistivity $\rho_{xy}$ and isothermal magnetizations at different angles $\theta$ are shown in Figs. S21(a) and S21(b), respectively. Inset in Fig. S21(a) shows the schematic measurement configuration. As increases $\theta$ from 0° to 90°, $\rho_{xy}$ is gradually reduced. In Fig. S21(c), it is clear that the DW Hall resistivity $\rho^{DW}_{xy}$ can be substantially influenced by varying $\theta$. As shown in Figs. S22(a)-S22(i), $\rho_{xy}$ at various rotation angles is fitted with the conventional two-band model, which reveals a clear deviation below 4 T, which signifies the presence of DW contribution as well as its evolution against $\theta$.

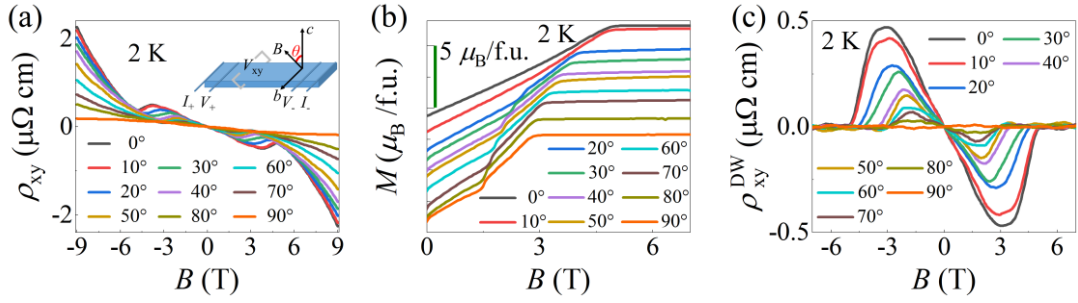

**Fig. S21.** Magnetic field dependent (a) Hall resistance $\rho_{xy}$, (b) isothermal magnetization and (c) DW Hall resistance $\rho^{DW}_{xy}$ at different angles and 2 K.

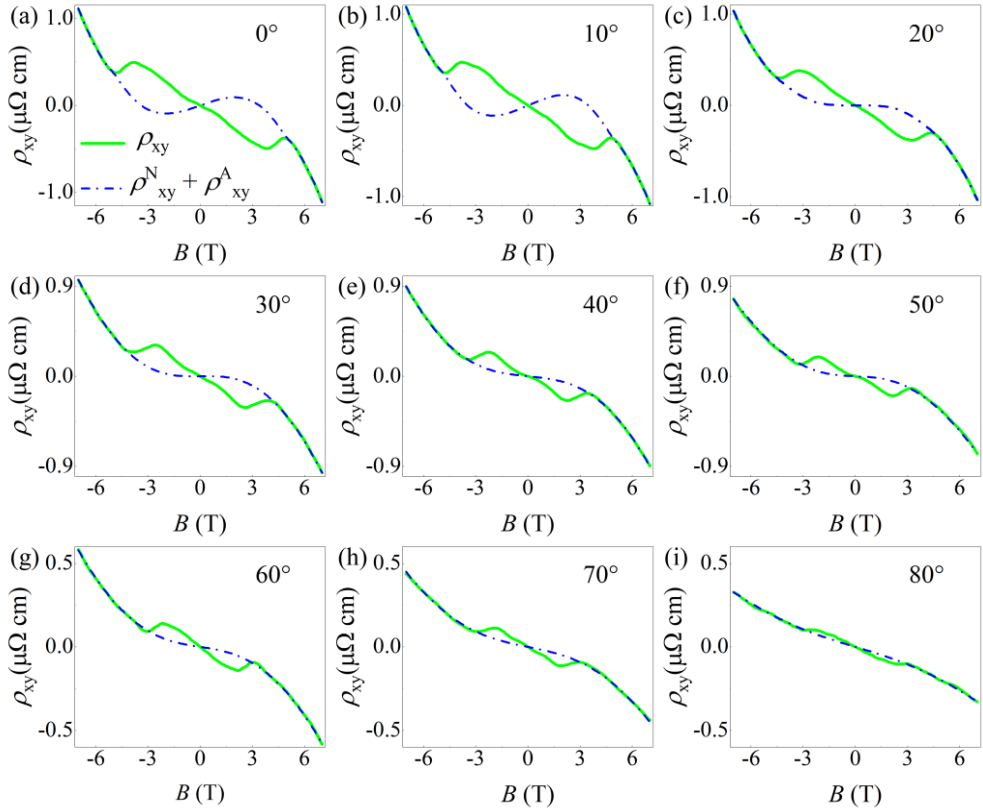

**Fig. S22.** (a)-(i) Hall resistivity (green markers) fitted to the conventional Hall contribution $\rho_{xy} = \rho^{N}_{xy} + \rho^{A}_{xy}$ at 0°- 80° (blue solid lines) and $B//c$-axis to $B \perp c$-axis. The clear deviation from the conventional Hall resistivity indicates the presence of an additional term.